\documentclass{aa}  

\usepackage{graphicx}
\usepackage{natbib}
\usepackage{txfonts}
\usepackage{lscape}
\usepackage{amstext}

\begin{document}

   \title{
Cygnus survey with the Giant Metrewave Radio Telescope at 325 and 610 MHz: the catalog}


   \author{P. Benaglia
          \inst{1,2}
          \and
          C. H. Ishwara-Chandra \inst{3}
          \and
          H. Intema\inst{4,5}
          \and 
          M. E. Colazo\inst{6}
          \and
          M. Gaikwad\inst{7}
          }

   \institute{Instituto Argentino de Radioastronom\'{\i}a, CONICET \& CICPBA, CC5 (1897) Villa Elisa, Prov. de Buenos Aires, Argentina\\
              \email{paula@iar.unlp.edu.ar}
         \and
             Facultad de Ciencias Astron\'{o}micas y Geof\'{\i}sicas, UNLP, Paseo del Bosque s/n, 1900, La Plata, Argentina
        \and
            National Centre for Radio Astrophysics (NCRA-TIFR), Pune, 411 007, India
        \and
            International Centre for Radio Astronomy Research, Curtin University, Bentley, WA 6102, Australia
        \and
            Leiden Observatory, Leiden University, Niels Bohrweg 2, 2333 CA Leiden, the Netherlands
        \and
            Comisi\'{o}n Nacional de Actividades Espaciales, Paseo Col\'{o}n 751 (1063) CABA, Argentina
        \and    
            Max-Planck-Institut für Radioastronomie, Auf dem Hügel 69, D-53121 Bonn, Germany
             }

   \date{Received xx, 2020; accepted YYY}

 
  \abstract
   {Observations at the radio continuum band below the gigahertz band are key when the nature and properties of nonthermal sources are investigated because their radio radiation is strongest at these frequencies. The low radio frequency range is therefore the best to spot possible counterparts to very high-energy (VHE) sources: relativistic particles of the same population are likely to be involved in radio and high-energy radiation processes. Some of these counterparts to VHE sources can be stellar sources. }
   {The Cygnus region in the northern sky is one of the richest in this type of sources that are potential counterparts to VHE sources.  We surveyed the central $\sim$15~sq deg of the Cygnus constellation at the 325 and 610~MHz bands with angular resolutions and sensitivities of $10''$ and $6''$, and 0.5 and 0.2 mJy~beam$^{-1}$, respectively.}
   {The data were collected during 172 hours in 2013 -- 2017, using the Giant Metrewave Radio Telescope (GMRT) with 32~MHz bandwidth, and were calibrated using the SPAM routines. The source extraction was carried out with the PyBDSF tool, followed by verification through visual inspection of every putative catalog candidate source in order to determine its reliability.}
   {In this first paper we present the catalog of sources, consisting of 1048 sources at 325~MHz and 2796 sources at 610~MHz. By cross-matching the sources from both frequencies with the objects of the SIMBAD database, we found possible counterparts for 143 of them. {Most of the sources from the 325-MHz catalog (993)} were detected at the 610~MHz band, and their spectral index $\alpha$ was computed adopting $S(\nu) \propto \nu^\alpha$. The maximum of the spectral index distribution is at $\alpha=-1$, { which is characteristic of nonthermal emitters and might indicate an extragalactic population.} 
  }  
   {}

   \keywords{Catalogs -- Radio continuum: general -- Open clusters and associations: individual: Cygnus OB2, OB8, OB9}

   \maketitle
%


\section{Introduction}
\label{sec:intro}

The first gamma-ray all-sky observations, obtained decades ago with the satellites COS-B \citep[][and references therein]{hermsen1977} and Compton \citep{hartman1999}, disclosed numerous sources with no counterpart at other wavelengths. These are hereafter called unidentified gamma-ray sources, or UNIDS.
Since then, a large number of multifrequency observations have been implemented to understand the nature of these sources \citep[e.g.,][]{paredes2008,massaro2013}. Despite significant improvement in the telescope capabilities in sensitivity and resolution, there still remain thousands of gamma-ray sources to be identified. 
For instance, the fourth catalog of the Fermi Large Area Telescope \citep[][more than 5000 sources]{Fermi4th} listed about one-third of the detected sources without any counterpart at lower energies. The sources that were detected with ground-based telescopes, at TeV energies, also present
problems in conclusive identification; in addition, the high uncertainty on their position precludes the correlation with individual objects (see, e.g., the High Energy Stereoscopic System, H.E.S.S., source catalog\footnote{https://www.mpi-hd.mpg.de/hfm/HESS/pages/home/sources/} and its identifications).

The identified gamma-ray sources are mostly active galactic nuclei, AGNs, and pulsars, supernova remnants, or high-mass X-ray binaries (HMXBs). These objects emit at radio wavelengths and are generally stronger at low radio frequencies ($<$ 1 GHz) as a result of the nature of the spectra of synchrotron radiation. In this part of the electromagnetic spectrum, major catalogs and surveys lack angular resolution or sensitivity to seek for singular counterparts of UNIDS \citep[e.g., the National Radio Astronomy Observatory Very Large Array Sky Survey, NVSS, $\sim45''$ and 1~mJy, or the Westerbork Northern Sky Survey, $\sim54''$ and 3~mJy; ][]{NVSS,WENSS}. 
Recently, other types of stellar sources have been proposed as possible gamma-ray emitters, and different scenarios were analyzed. In addition to the well-studied microquasars \citep{romero2003}, colliding-wind binaries \citep{benagliaromero2003}, Herbig Haro objects, young stellar objects (YSO), \citep{valenti2010,araudo2007,rodriguez2019}, and stellar bow shocks \citep{benaglia2010,delvalle2018,delpalacio2018} are capable of producing
gamma-rays. A signature of high-energy emission is nonthermal radio emission because particles from the same population are likely to be involved in processes at both energy ranges, at the radio through synchrotron process, and at VHE emission through inverse-Compton scattering. Moreover, the determination of counterparts of gamma-ray sources through radio observations in star-forming regions will help to clarify the role of young stars and collective wind effects in the acceleration of galactic cosmic rays \citep[e.g.,][]{romero2008}.

Various high-energy sources have been detected in the northern-sky Cygnus rift, a large region with star-formation activity that is one of the richest and most crowded in stellar objects in the Galaxy. Many thousand sources are cataloged in the literature in this region, and more than half are stars. The high absorption in the line of sight, however, prevents accurate mapping of the stellar population at the optical and IR ranges. 
Low-frequency (centimeter wavelengths) observations are the only way to probe nonthermal radio emission, and this emission travels practically unabsorbed; observing facilities that provide high angular resolution and sensitivity are crucial. In this line, the Giant Metrewave Radio Telescope (GMRT) is ideal for sampling the sky to search for emission of stellar sources: it operates between 150 and 1400~MHz, with baselines along 25~km that allow images with an angular resolution of a few arcseconds \citep{swarup1991}.

We carried  out a survey of the center of the Cygnus rift with the GMRT by means of continuum observations at two bands (325 MHz and 610 MHz) to investigate the nonthermal emission of various types of sources that lie in this rich field and are potential counterparts of UNIDS. With two frequencies, we were also able to obtain spectral information that might help to categorize certain classes and emission mechanisms on the basis of the spectral index. 
We present the  source catalog at each band here, along with spectral index information when possible. In Section 2 we present the main characteristics of the Cygnus region and precedent studies at low-frequency radio continuum; in Sect. 3 and 4 we describe how the observations were carried out, and the processes attached to the data reduction to obtain the final images. Section 5 explains the data analysis we performed on the images and how the sources were extracted.  Section 6 contains the findings related to spectral indices for the sources we detected at the two observing bands. In Section 7 we discuss the main properties of the catalog. Results of the search for counterparts are given in Sect. 8, and we conclude by mentioning related studies and prospects in the last section.

\section{Cygnus region and background of the radio observations}
\label{sec:cyg}

\begin{figure}
   \centering
   \includegraphics[width=\columnwidth]{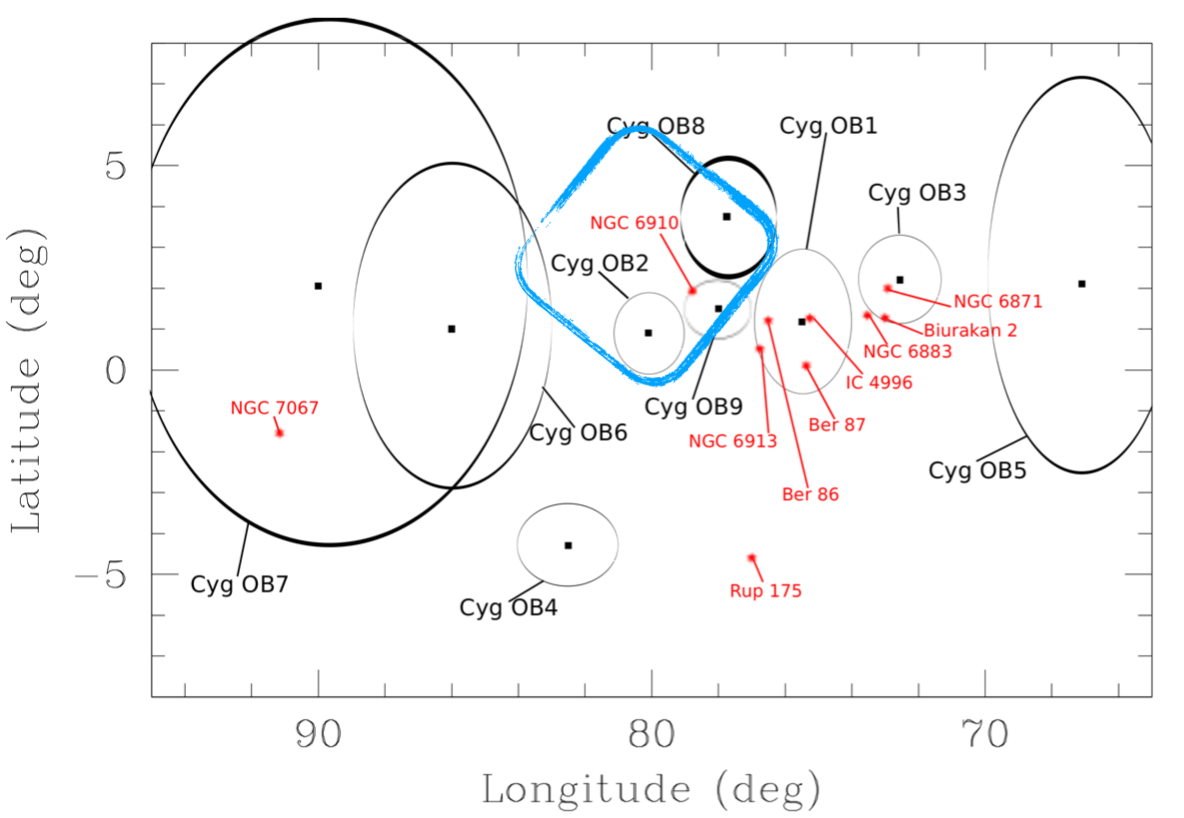}
      \caption{Observed area of the Cygnus constellation marked with a blue contour box, over the stellar associations and bright clusters from \cite{mahy2013}.}
         \label{fig:Mahy}
   \end{figure}

The Cygnus rift is a large area at northern declination that is obscured by the dust of molecular clouds. It spans from $65^\circ \leq l \leq 95^\circ$, $-8^\circ \leq b \leq +8^\circ$ at a distance up to 2.5~kpc; see \cite{reipurthcyg} for a comprehensive review.
As portrayed in Fig.~\ref{fig:Mahy} \citep{mahy2013}, it encompasses nine OB associations and several bright open clusters, with signs of recent star formation. One of the youngest associations is Cyg\,OB2: it is also the richest association, with more than one hundred O stars and thousands of B stars, as reported by \cite{knodl2000}. Next to OB2, Cyg\,OB8 and Cyg\,OB9 present hundreds of hot stars.

The main goal of the project was to relate nonthermal radio sources with stellar objects, that is,~stars at different evolutionary stages:~we circumscribed the region under study to the the associations Cyg\,OB2, OB8 and OB9; its extension is displayed in Fig.~\ref{fig:Mahy} in Galactic coordinates, related to the Cygnus constellation. It covers $\sim$15 sq deg. The associations Cyg\,OB8  and OB9 are not surveyed in full because they are adjacent to strong and/or large sources (like the cases of Cyg\,X--1 and Cyg\,A), which might introduce problems related to imaging of a highly dynamic range.

Below the Jansky threshold, the Cygnus area was observed as part of Galactic plane surveys with the Very Large Array\footnote{https://science.nrao.edu/facilities/vla} by \cite{garwood1988} at 1.4~GHz continuum, at $b=0^\circ$, a resolution up to $4''$, and completely to a peak flux density of about 30~mJy. The results were complemented by those of \cite{zoone1990} for $|b| < 0.8^\circ$ , which provided angular resolution and flux limit alike. With the Texas Interferometer, \cite{douglas1996} imaged the area at the arcsecond scale above flux densities of 0.25--0.4~Jy. \cite{taylor1996} carried out Westerbork Synthesis Radio Telescope observations along the Galactic plane and for $|b| < 1.6^\circ$, at an angular resolution of $\sim1'$. They detected sources brighter than 10~mJy~beam$^{-1}$.

In particular, \cite{setia2003} published the Westerbork Synthesis Radio Telescope 1400 and 325~MHz continuum survey of Cyg~OB2, which attained angular resolutions of $13''$ and $55''$ and 5 $\sigma$ flux density limits of $\sim$2~mJy and $\sim$10--15~mJy, respectively. In an observed area of $2^\circ \times 2^\circ$, the authors detected 210 discrete sources, 98 of them at both frequencies. They also detected 28 resolved sources.

The observations presented here were performed at two bands with the GMRT, centered at 325 and 610~MHz. This allowed us to map the continuum radio emission at arcsecond resolution and below the mJy sensitivity level. Some information about the observations has been given in \cite{ishwar2019}.

\section{Observations}

\begin{figure}
   \centering
   \includegraphics[width=8cm]{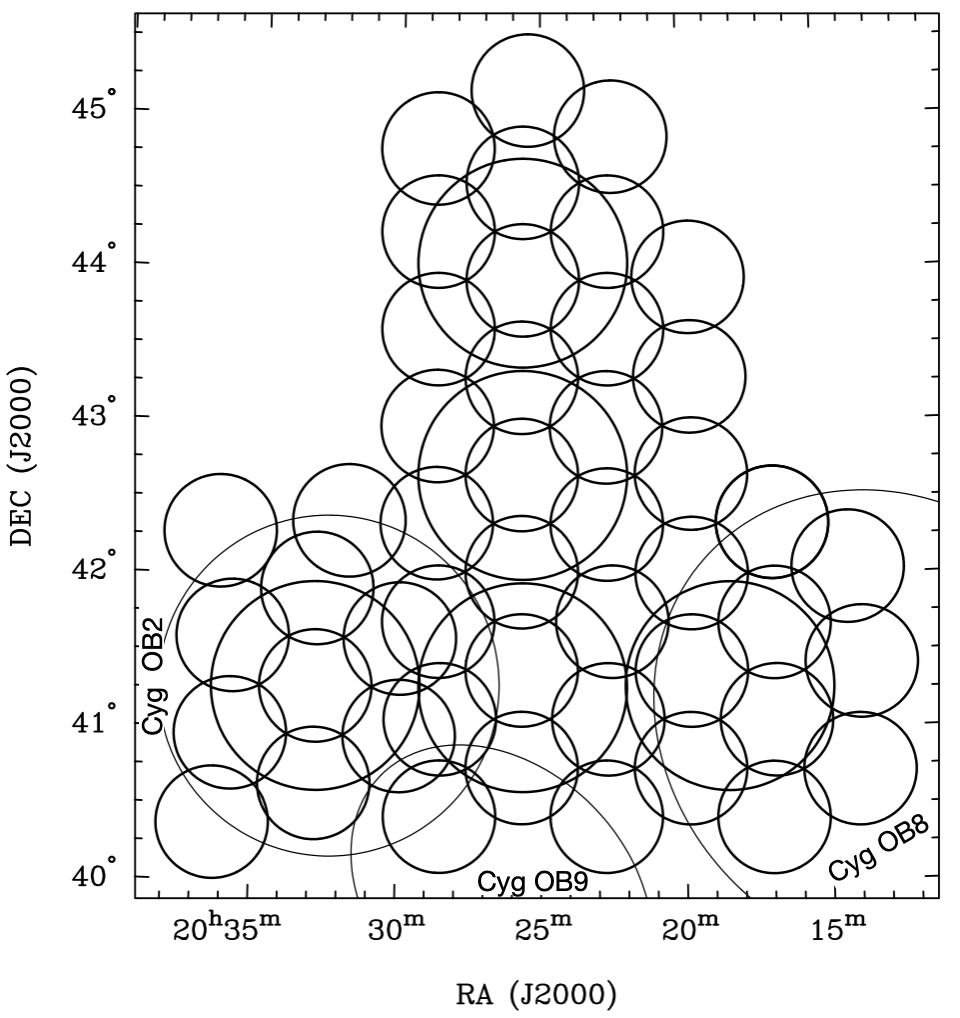}
      \caption{Disposition of the pointings at 325~MHz (larger circles) and 610~MHz (smaller circles), showing the observed FoV half-power beam widths.  The extent of the stellar associations Cyg\,OB2, OB8, and OB9 \citep{uyaniker2001} is shown using thinner lines.}
         \label{fig:fovs}
   \end{figure}

The observed region marked in Fig.~\ref{fig:Mahy} is displayed in equatorial coordinates in Fig.~\ref{fig:fovs}. The half-power beam widths of the GMRT fields of view (FoVs) are 81$\pm$4$'$ at 325~MHz and 43$\pm$3$'$ at 610~MHz\footnote{GMRT Observer’s Manual; www.ncra.tifr.res.in/ncra/gmrt/gmrt-users/observing-help/manual\_7jul15.pdf}.

To cover the desired observing area, we needed to point at 5\ FoVs of 325~MHz  and 47 FoVs of 610~MHz. Some observations consisted of bad data, so that we repeated them with new observations (20~h). The layout of the pointings was chosen to yield a uniform noise while minimizing the number of them (i.e., the observing time). Figure~\ref{fig:fovs} shows the disposition of the pointings and the FoVs at both observing bands. The project was divided into four observing campaigns, scheduled from November 2013 to September 2017. Table~\ref{table:obscampaigns} lists the GMRT campaign ID, the allocated time, and the year(s) of completion. 

Details of the targeted areas and observing parameters are given in Table~\ref{table:obsruns}: the name of the FoV, the corresponding campaign ID, the exact observation date, the position of the pointing center, the time on the FoVs (t.o.s.), the band, and the calibrators, ordered by band and by right ascension. The observations were carried out using the total intensity mode and a bandwidth of 32 MHz and 256 spectral channels to minimize the effect of bandwidth smearing. Flux calibrators were observed at the start and end of the each run for flux and bandpass calibration. A phase calibrator was observed for 5 minutes after a scan of 30 minutes on the target to calibrate the phases and any slow variations of the telescope gain.

%
\begin{table}
\caption{Observing campaigns basic information.}
\label{table:obscampaigns}
\centering             
\begin{tabular}{l c c c}       
\hline\hline              
\# & Campaing ID & Time (h) & Obs. dates \\  
\hline  
   1 & 25\_026 & 12 & 2013 \\
   2 & 27\_036 & 40 & 2014--2015\\
   3 & 28\_081 & 40 & 2015 \\
   4 & 30\_027 & 60 & 2016--2017\\
\hline  
\end{tabular}
\end{table}
%

%
\begin{table*}
\caption{Observing runs and fields of view information.}        
\label{table:obsruns}      
\centering          
\begin{tabular}{l l r c c r r r}  
\hline\hline     
 Field of view & Campaign & Observing & \multicolumn{2}{c}{Pointing center (J2000)} & t.o.s. & Band & Calibrators\\
      \,\,\,\,\,\,\,\,name    &\,\,\,\,\,\,\,\, ID       &   dates        & RA (h,m,s) & Dec (deg,$'$,$''$)& (min) & (MHz) & \\
\hline                    
FoV325.1 & 27\_036 & 7/2,26/9/15 & 20 18 26 & 41 16 50 & 488 & 325 & 3C48,2052+365 \\ 
FoV325.2 & 27\_036 & 26/10/14 & 20 25 38 & 41 16 50 & 304 & 325 & 3C48,2052+365 \\ 
FoV325.3 & 27\_036 & 27/10/14 & 20 25 38 & 42 39 50 & 296 & 325 & 3C147,2052+365 \\ 
FoV325.4 & 27\_036 & 6/2/15 & 20 25 38 & 44 02 50 & 283 & 325 & 3C48,2052+365 \\ 
FoV325.5 & 25\_026 & 4/11/13 & 20 32 50 & 41 16 50 & 523 & 325 & 3C48,2038+513 \\ 
\hline %
FoV610.1 & 28\_081 & 18/6/15 & 20 13 50 & 41 21 15 & 79 & 610 & 3C48,2052+365 \\ 
FoV610.2 & 28\_081 & 18/6/15 & 20 13 60 & 40 43 10 & 79 & 610 & 3C48,2052+365 \\ 
FoV610.3 & 30\_027 & 11,21/8/16 & 20 14 13 & 42 02 30 & 118 & 610 & 3C48,3C286,2052+365 \\ 
FoV610.4 & 28\_081 & 18/6/15 & 20 16 50 & 41 03 15 & 130 & 610 & 3C48,2052+365 \\ 
FoV610.5 & 28\_081 & 25/7/15 & 20 16 50 & 41 41 20 & 119 & 610 & 3C48,2052+365 \\ 
FoV610.6 & 30\_027 & 2/9/17 & 20 16 50 & 42 20 30 & 69 & 610 & 3C48,2052+365 \\ 
FoV610.7 & 28\_081 & 25/7/15 & 20 17 00 & 40 25 05 & 71 & 610 & 3C48,2052+365 \\ 
FoV610.8 & 30\_027 & 17/7,8/8/16 & 20 19 40 & 42 40 01 & 147 & 610 & 3C48,2052+365 \\ 
FoV610.9 & 30\_027 & 17/7,8/8/16 & 20 19 40 & 43 18 01 & 179 & 610 & 3C48,2052+365 \\ 
FoV610.10 & 30\_027 & 17/7/16 & 20 19 40 & 43 56 56 & 78 & 610 & 3C48,2052+365 \\ 
FoV610.11 & 30\_027 & 1/7/16 & 20 19 42 & 42 01 08 & 94 & 610 & 3C286,2052+365 \\ 
FoV610.12 & 28\_081 & 25/7/15 & 20 19 45 & 41 22 60 & 119 & 610 & 3C48,2052+365 \\ 
FoV610.13 & 28\_081 & 15/8/15 & 20 19 50 & 40 44 50 & 119 & 610 & 3C48,2052+365 \\ 
FoV610.14 & 30\_027 & 15/7/16 & 20 22 25 & 44 52 09 & 64 & 610 & 3C48,2052+365 \\ 
FoV610.15 & 28\_081 & 16/8/15 & 20 22 30 & 41 42 30 & 119 & 610 & 3C48,2052+365 \\ 
FoV610.16 & 30\_027 & 2/9/17 & 20 22 34 & 44 15 01 & 59 & 610 & 3C48,2052+365 \\ 
FoV610.17 & 30\_027 & 2/9/17 & 20 22 36 & 43 36 53 & 59 & 610 & 3C48,2052+365 \\ 
FoV610.18 & 28\_081 & 15/8/15 & 20 22 40 & 41 04 20 & 107 & 610 & 3C48,2052+365 \\ 
FoV610.19 & 30\_027 & 30/6,23/7/16 & 20 22 40 & 42 20 30 & 168 & 610 & 3C48,3C286,2052+365 \\ 
FoV610.20 & 30\_027 & 30/6,23/7/16 & 20 22 40 & 42 58 30 & 231 & 610 & 3C48,3C286,2052+365 \\ 
FoV610.21 & 28\_081 & 15/8/15 & 20 22 45 & 40 26 15 & 62 & 610 & 3C48,2052+365 \\ 
FoV610.22 & 30\_027 & 15/7/16 & 20 25 27 & 45 10 21 & 71 & 610 & 3C48,2052+365 \\ 
FoV610.23 & 30\_027 & 2/9/17 & 20 25 38 & 43 56 06 & 76 & 610 & 3C48,2052+365 \\ 
FoV610.24 & 30\_027 & 2/9/17 & 20 25 38 & 44 34 14 & 70 & 610 & 3C48,2052+365 \\ 
FoV610.25 & 28\_081 & 16/8/15 & 20 25 40 & 40 45 25 & 119 & 610 & 3C48,2052+365 \\ 
FoV610.26 & 28\_081 & 17/8/15 & 20 25 40 & 41 23 35 & 142 & 610 & 3C48,2052+365 \\ 
FoV610.27 & 30\_027 & 14/7/16 & 20 25 40 & 42 01 60 & 89 & 610 & 3C48,2052+365 \\ 
FoV610.28 & 30\_027 & 30/6,23/7/16 & 20 25 40 & 42 40 00 & 168 & 610 & 3C48,3C286,2052+365 \\ 
FoV610.29 & 30\_027 & 1,2,24/7/16 & 20 25 40 & 43 18 00 & 243 & 610 & 3C48,3C286,2052+365 \\ 
FoV610.30 & 28\_081 & 17/8/15 & 20 28 30 & 40 26 15 & 79 & 610 & 3C48,2052+365 \\ 
FoV610.31 & 28\_081 & 16/8/15 & 20 28 30 & 41 04 20 & 71 & 610 & 3C48,2052+365 \\ 
FoV610.32 & 28\_081 & 17/8/15 & 20 28 35 & 41 42 30 & 79 & 610 & 3C48,2052+365 \\ 
FoV610.33 & 30\_027 & 1,2,24/7/16 & 20 28 40 & 42 21 00 & 232 & 610 & 3C48,3C286,2052+365 \\ 
FoV610.34 & 30\_027 & 1,2,24/7/16 & 20 28 40 & 42 59 00 & 230 & 610 & 3C48,3C286,2052+365 \\ 
FoV610.35 & 30\_027 & 1/7/16 & 20 28 40 & 43 36 53 & 94 & 610 & 3C286, 2052+365 \\ 
FoV610.36 & 30\_027 & 1/7/16 & 20 28 42 & 44 15 01 & 94 & 610 & 3C286, 2052+365 \\ 
FoV610.37 & 30\_027 & 15/7/16 & 20 28 44 & 44 47 32 & 71 & 610 & 3C48, 2052+365 \\ 
FoV610.38 & 27\_036 & 28,29/11/14 & 20 29 55 & 40 57 38 & 120 & 610 & 3C48,2052+365 \\ 
FoV610.39 & 27\_036 & 28,29/11/14 & 20 29 55 & 41 35 45 & 120 & 610 & 3C48,2052+365 \\ 
FoV610.40 & 30\_027 & 17/7,8/8/16 & 20 31 45 & 42 21 39 & 145 & 610 & 3C48,2052+365 \\ 
FoV610.41 & 27\_036 & 28,29/11/14 & 20 32 50 & 40 38 42 & 120 & 610 & 3C48,2052+365 \\ 
FoV610.42 & 30\_027 & 14/7,8/8/16 & 20 32 50 & 41 16 50 & 165 & 610 & 3C48,2052+365 \\ 
FoV610.43 & 27\_036 & 28,29/11/14 & 20 32 50 & 41 54 58 & 120 & 610 & 3C48,2052+365 \\ 
FoV610.44 & 27\_036 & 28,29/11/14 & 20 35 45 & 40 57 38 & 105 & 610 & 3C48,2052+365 \\ 
FoV610.45 & 27\_036 & 28,29/11/14 & 20 35 45 & 41 35 45 & 120 & 610 & 3C48,2052+365 \\ 
FoV610.46 & 30\_027 & 8,11,21/8/16 & 20 36 17 & 40 22 30 & 177 & 610 & 3C48,3C286,2052+365 \\ 
FoV610.47 & 30\_027 & 11,21/8/16 & 20 36 17 & 42 16 30 & 118 & 610 & 3C48,3C286,2052+365 \\ 
\hline                  
\end{tabular}
\end{table*}

\section{Data reduction} 
\label{datared}

The five pointings at 325~MHz were processed uniformly, using the SPAM routines \citep{IntemaSPAM},  which is a {python package based on the Astronomical Image Processing System (AIPS)} for nearly automatic analysis of GMRT data below 1~GHz. Bad data and RFI were initially flagged at the full spectral resolution of 256 channels.  
The flux was calibrated using the primary calibrators {3C286, 3C48, and 3C147} and the scale by \citet{2012MNRAS.423L..30S} for low radio frequencies. 
The SPAM pipeline then converted the precalibrated visibility data to a final image, which includes several rounds  of self-calibration and flagging iteratively, and wide-field imaging to correct for noncoplanarity. In the self-calibration, ionospheric phase corrections were computed for several directions within the FoV for direction-dependent corrections on the integration timescales.
The self-calibration procedure was followed using default parameters of SPAM, with initial cycles in phase with long intervals and a last run with the solution interval of the visibility integration time; in our case, 16.9~seconds. Toward the end of the loop, one round of amplitude and phase self-calibration was carried out.
During imaging, a moderately uniform weighting scheme (robust=$-1$ in AIPS) was used, but no multiscale cleaning options were incorporated. 
Primary-beam corrections were applied using the GMRT specific parameters (GMRT Observer's Manual).

Because the target is in the Galactic plane, the $T_{\rm sys}$ correction for excess background was applied using the 408 MHz all-sky map \citep{1982A&AS...47....1H} during the calibration as part of the pipeline.
The correction factor varied from 1.7 to 3.6, with the highest correction factor near the  Galactic plane and a lower correction factor away from the plane. The FoVs were combined in a mosaic with weights as inverse of the variance. 

The data analysis of the 47 FoVs at 610~MHz was also carried out using the SPAM pipeline, similar to the 325 MHz data. The $T_{\rm sys}$ correction for the excess background at 610 MHz ranged from 1.22 to 1.76. Fields of view FoV610.21 and FoV610.30 were noisier than the rest, probably because of strong extended emission from the Galactic plane and bright sources in the fields.  

\begin{figure*}
   \centering
   \includegraphics[width=18cm]{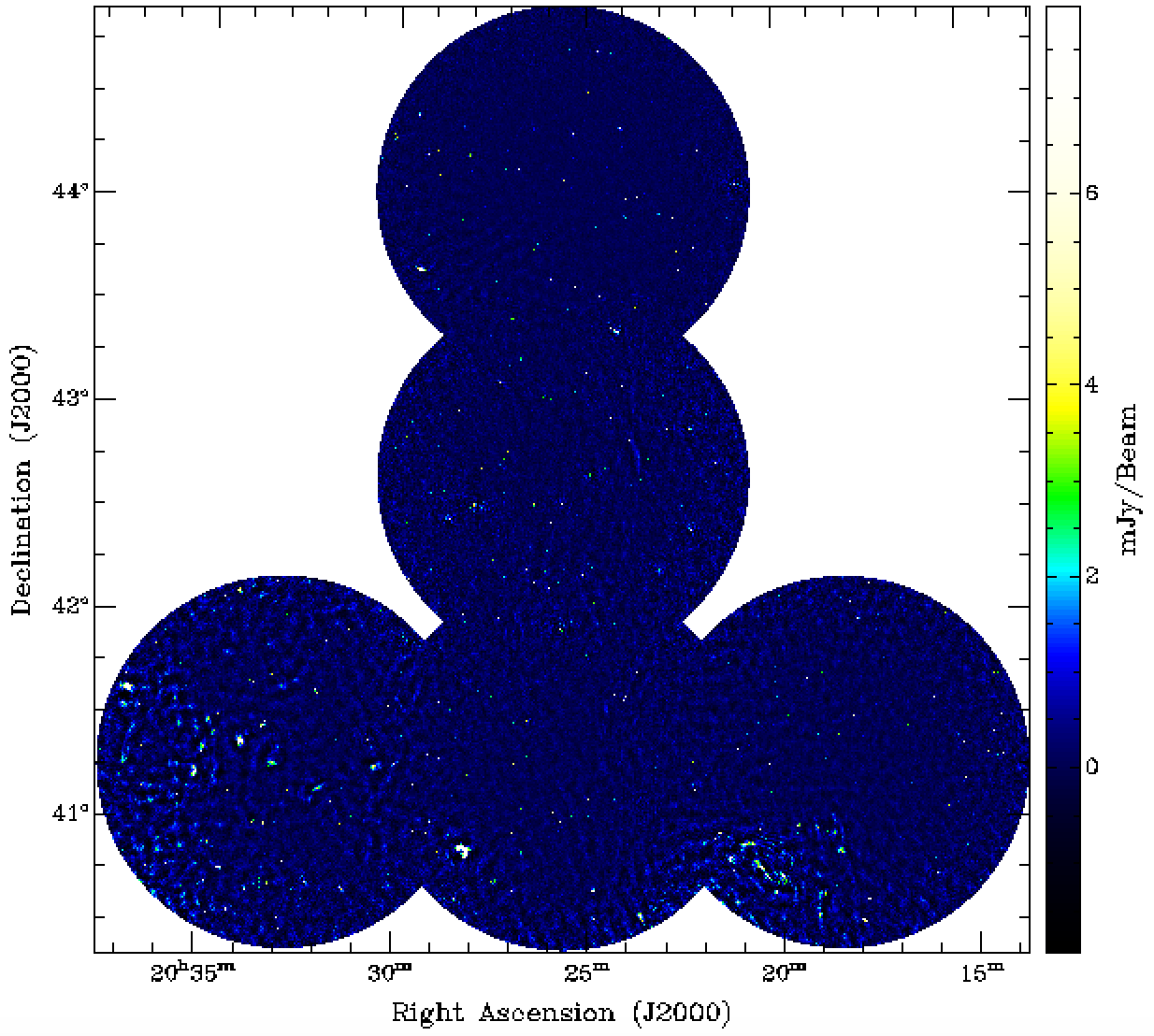}
      \caption{GMRT 325 MHz continuum image of the observed field. The synthesized beam is $10'' \times 10''$, and the average rms is 0.5 mJy beam$^{-1}$. The  full range of the flux density values is $-$9.2,$+$812.5~mJy~beam$^{-1}$. The interval shown is ($-$2,$+$8)~mJy~beam$^{-1}$ to outline the weaker features.}
         \label{fig:mosaic325}
   \end{figure*}

\begin{figure*}
   \centering
   \includegraphics[width=18cm]{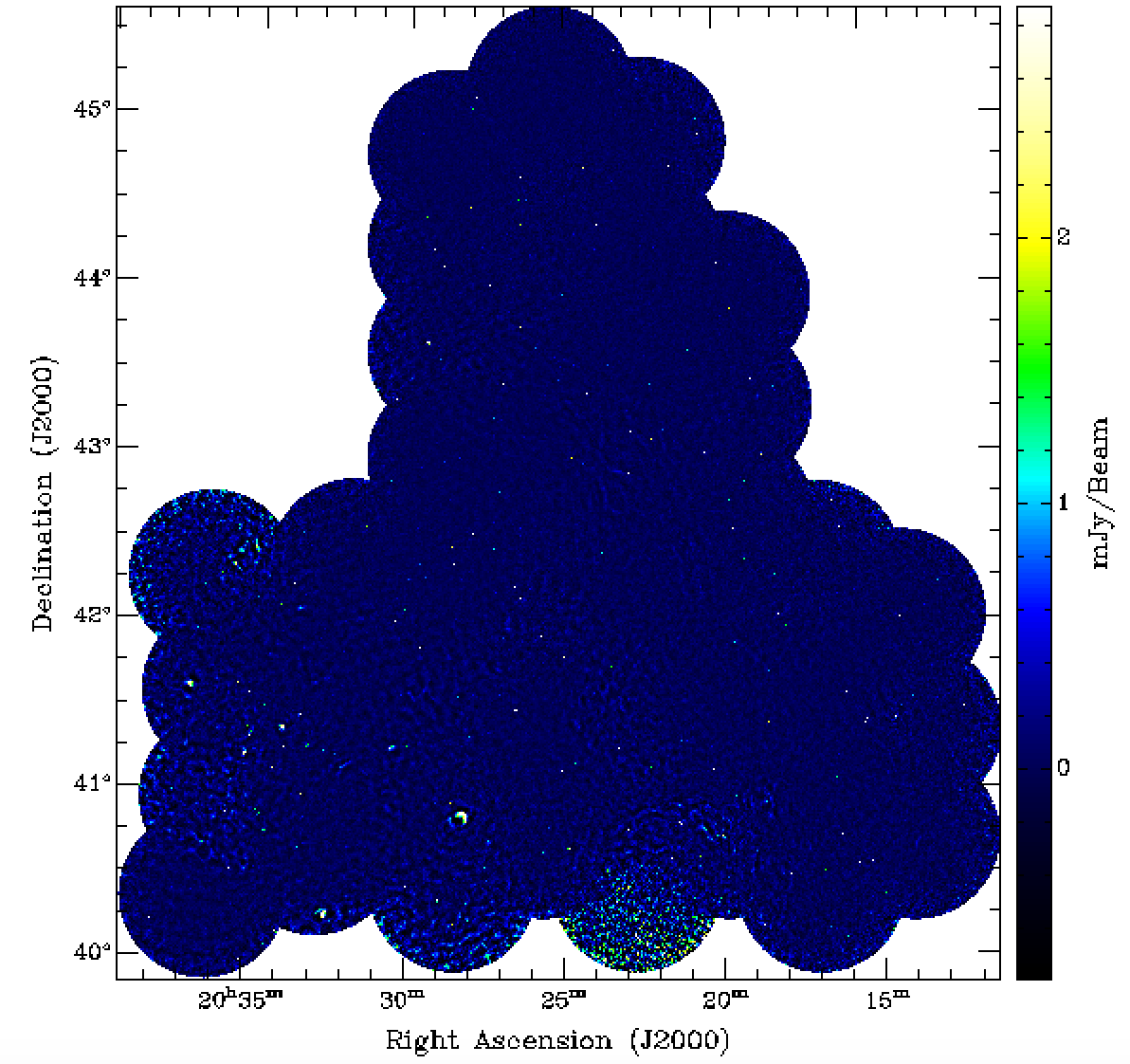}
      \caption{GMRT 610 MHz continuum image of the observed field. The synthesized beam is $6'' \times 6''$, and the average rms is 0.2 mJy beam$^{-1}$. The full range of the flux density values is $-$6.3,$+$928.3~mJy~beam$^{-1}$. The interval shown is ($-$0.8,$+$2.9)~mJy~beam$^{-1}$ to outline the weaker features.}
         \label{fig:mosaic610}
   \end{figure*}

The final mosaics presented an average rms of 0.5~mJy~beam$^{-1}$ and 0.2~mJy~beam$^{-1}$ at 325 and 610~MHz, respectively, although locally, the values strongly depend mainly on the extended and/or diffuse emission. The synthesized beams attained were $10'' \times 10''$, and $6'' \times 6''$, and the mosaic image sizes were ($6487 \times 6573$), and ($12580 \times 13837$) pixels, respectively. The final images are presented in Figs.\,\ref{fig:mosaic325}  and  \ref{fig:mosaic610}.

A series of factors affect the accuracy of low-frequency radio flux scales in different ways. This in turn affects the uncertainties. In their observations with the GMRT, \citet{Chandra2004}, for instance, discussed a flux-scale uncertainty at 325 and 610 MHz of a few percent. In addition, when the target fields lie in the Galactic plane, as the field we present, the fact that $T_{\rm sys}$ can be significantly higher because the sky temperature is higher than toward the calibrator imposes an additional correction factor; although the SPAM pipeline estimates this factor, it is based in some assumptions and extrapolations that may introduce some inaccuracy. We used different flux calibrators, the primary-beam model is not perfect, and mosaicking different pointings into the final images we used for source extraction can all affect the flux scale. When all of this is taken into account, a very conservative approach is to adopt flux density uncertainties of 10\%.

\section{Source extraction}\label{sec:srcextraction}

   \begin{figure}[!h]
   \centering
   \includegraphics[width=\columnwidth]{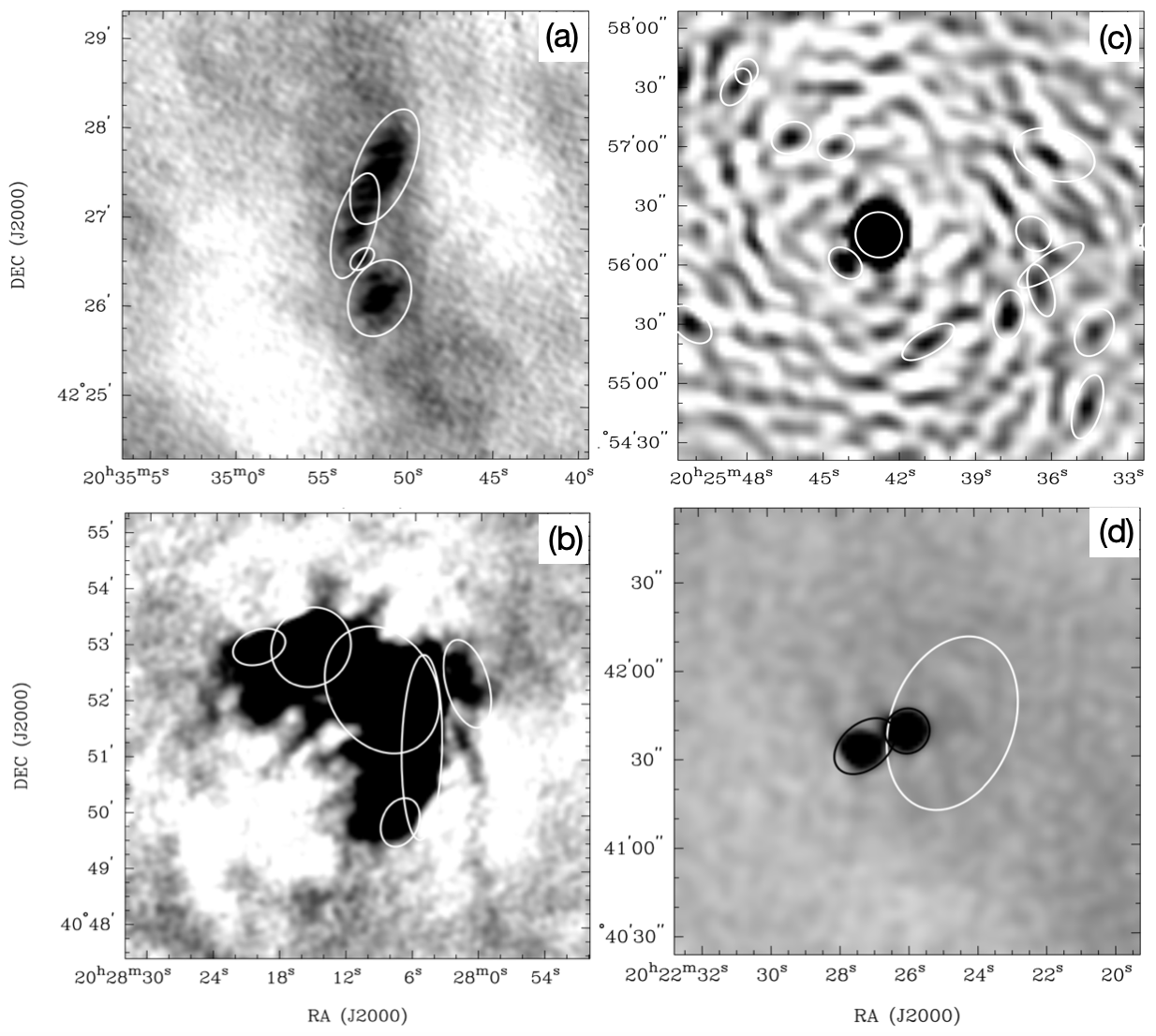}
      \caption{Examples of discarded fits for the four cases given in Sect.\,\ref{sec:srcextraction},  represented as white-line ellipses. (a): Filaments and/or diffuse emission at 610~MHz. (b): Strong or large sources that are ill represented by a combination of fits at 325~MHz. (c): Reduction artifacts at 325~MHz. (d): Fits of emission similar to the noise at 610~MHz; in this last case, the fits accepted as good fits are shown with black-line ellipses.}
         \label{fig:discarded}
   \end{figure}
When the mosaics at the two observing bands were built, we searched for errors in the astrometry,  first, by taking point sources in the entire two images into consideration. We did not find significant errors at the smaller pixel scale, which means that the accuracy was better than 1.5 arcseconds.  We also searched the 610~MHz image and point sources with well-determined optical positions, that is, Wolf-Rayet and O-type stars. For the cases with radio emission at or near the position of these stars (13 in total), the differences in coordinates were in the  range 0.1 -- 3.1$"$, and the standard deviation was 1.34$"$ 
\citep[see][for a study of the massive early-type stars detected in the current databases]{cygnusstars2020}.

To survey the emission in the 325 and 610~MHz images, we applied the Python blob detector and source finder (PyBDSF\footnote{http://www.astron.nl/citt/pybdsf/}). The tool can be used to find islands of emission in radio interferometry images, decompose them into Gaussian functions, and finally gather them into individual source fits. We chose a signal-to-noise ratio of 7 as the lower limit for a source or fit to be accepted, and proceeded in the same way as \cite{benaglia2019}, where this proved successful. A similar detection threshold was used for the CORNISH catalog \citep{CORNISH}. The routine includes the determination of the rms in the image and the production of a corresponding rms image. 


 The running of PyBDSF over the 325 and 610~MHz image mosaics resulted in 1230 and 3023 sources, respectively. After a thorough visual inspection of each source, we kept 1048 (85.2\%) sources from the 325~MHz image and 2796 (92.5\%) sources from the 610~MHz image. 

The group of accepted entries consisted of discrete (unresolved) objects that were represented by one fitted source of the size of the synthesized beam, and of resolved objects. Some resolved objects were represented by a fitted source that was larger than the synthesized beam, while others were described by a combination of fitted sources.

We rejected fits to filaments and (part of) diffuse emission (3.5\% at 325~MHz and 1.5\% at 610~MHz, see Fig.\,\ref{fig:discarded}-a) and the fit combinations that corresponded to strong and/or large objects with ill representations (1.7\% at 325~MHz and 2.0\% at 610~MHz; Fig.\,\ref{fig:discarded}-b). We also discarded either objects with reduction artifacts that precluded a proper fit (including end-of-field objects: 2.5\% at 325~MHz and 8.2\% at 610~MHz, Fig.\,\ref{fig:discarded}-c) and fitted sources similar to surrounding noise (1.5\% at 325~MHz and 1.3\% at 610~MHz, see Fig.\,\ref{fig:discarded}-d). 

Overall, the 610~MHz emission was better imaged by the SPAM pipelines than the emission at the lower frequency band. The 325~MHz mosaic presented a higher percentage of extended emission-fitting problems. The largest detectable structure is 32$'$ at the 325~MHz band and 17$'$ at the 610~MHz band (GMRT User's Manual): the data presented here are biased against structures that are larger than that. The selection of the robust weighting of $-1$, a compromise between high angular resolution and signal-to-noise ratio, outlined discrete sources over diffuse emission. 

The attained rms at the two bands, in addition to the intrinsic values contributed by the stellar sources under observation, are indeed a function of the time-on-source (here, time-on-fields). To determine the completeness of the sources detected above 7$\sigma$ (7\,rms), we also ran the PyBDSF routines using a detection threshold of 5$\sigma$. We found 1721 fitted sources at 325~MHz and 5015 at 610~MHz, which is well above the numbers found for the first run. 
When we visually inspected several faint sources, they appeared like noise peaks, which prompted us to use 7$\sigma,$ with which we found significantly fewer spurious sources. 
The incompleteness of our catalog is mainly quantified by the reasons given above regarding the fits.

The final lists of the accepted objects are reunited in our catalog.
The cataloged sources are given in Tables\,\ref{sources3} and \ref{sources6}, named consecutively (column~1) by increasing right ascension; only sample items are shown.
We  tagged the sources with the label ``BIC'' followed by the observing frequency in MHz, and then a correlative number, based on their order. For each source, we list the coordinates $RA$, $Dec$ (J2000) of the fit (columns 2 and 3), the integrated flux (column 4),  the peak flux (column 5) and the fitted major axis, minor axis, and position angle ($\theta_1$, $\theta_2$ , and $PA$, columns 6 to 8), which represent the source size and orientation after deconvolution, all with their corresponding errors {as reported by PyBDSF}. 
The full tables\,\ref{sources3} and \ref{sources6} are provided as online material.

\begin{table*}
\caption{Detected sources at 325~MHz above the 7$\sigma$ level (first entries); the full table with 1048 records is available as online material.}
\centering
\begin{tabular}{l@{~~~}r@{~~~}r@{~~~}r@{~~~}r@{~~~}r@{~~~}r@{~~~}r} 
\hline\hline    
ID & $RA_{\rm J2000}$ & $Dec_{\rm J2000}$ & Total flux & Peak flux & $\theta_1$ & $\theta_2$ & $PA$ \\ 
   & (h,m,s) & (deg,$'$,$''$) & (mJy)  & (mJy)  &  ($''$)  & ($''$)  & (\degr) \\
\hline
BIC325-0001 & 20:13:54.55$\pm$0.049 & 41:32:32.01$\pm$0.88 &  4.7$\pm$1.00 &  2.8$\pm$0.40 & 13.9$\pm$2.10 & 13.9$\pm$2.10 & 16.0$\pm$46.54\\
BIC325-0002 & 20:13:56.45$\pm$0.025 & 41:33:06.07$\pm$0.30 & 16.7$\pm$1.19 &  7.7$\pm$0.39 & 17.2$\pm$0.97 & 17.2$\pm$0.97 & 56.4$\pm$6.23\\
BIC325-0003 & 20:14:04.78$\pm$0.024 & 40:56:14.47$\pm$0.27 &  5.1$\pm$0.63 &  5.0$\pm$0.36 & 10.8$\pm$0.83 & 10.8$\pm$0.83 & 93.3$\pm$21.93\\
BIC325-0004 & 20:14:07.82$\pm$0.037 & 41:02:47.35$\pm$0.61 &  3.4$\pm$0.69 &  2.8$\pm$0.36 & 11.2$\pm$1.44 & 11.2$\pm$1.44 &  7.9$\pm$87.95\\
BIC325-0005 & 20:14:17.56$\pm$0.003 & 41:29:24.16$\pm$0.03 & 80.9$\pm$0.83 & 58.9$\pm$0.38 & 12.9$\pm$0.09 & 12.9$\pm$0.09 & 77.2$\pm$179.64\\
BIC325-0006 & 20:14:18.62$\pm$0.007 & 41:18:10.88$\pm$0.07 & 43.3$\pm$0.97 & 25.0$\pm$0.38 & 14.5$\pm$0.24 & 14.5$\pm$0.24 & 74.8$\pm$1.70\\
BIC325-0007 & 20:14:19.17$\pm$0.003 & 41:29:36.95$\pm$0.04 & 54.7$\pm$0.79 & 42.2$\pm$0.38 & 12.0$\pm$0.11 & 12.0$\pm$0.11 & 73.0$\pm$1.59\\
BIC325-0008 & 20:14:24.35$\pm$0.012 & 41:41:32.07$\pm$0.36 & 24.3$\pm$1.48 & 10.4$\pm$0.38 & 18.5$\pm$0.91 & 18.5$\pm$0.91 &  110.5$\pm$3.29\\
BIC325-0009 & 20:14:27.13$\pm$0.036 & 41:17:52.59$\pm$0.36 &  6.5$\pm$0.85 &  4.4$\pm$0.37 & 13.7$\pm$1.28 & 13.7$\pm$1.28 & 76.4$\pm$14.16\\
BIC325-0010 & 20:14:29.42$\pm$0.014 & 41:41:32.46$\pm$0.13 & 24.1$\pm$1.04 & 14.0$\pm$0.41 & 15.4$\pm$0.51 & 15.4$\pm$0.51 & 81.4$\pm$2.61\\
\hline
\end{tabular}
\tablefoot{$\theta_1$, $\theta_2$, and $PA$ represent the elliptic source size and orientation, and correspond to the major axis, the minor axis, and the position angle of the fit by the PyBDSM routines.}
\label{sources3}
\end{table*}

\begin{table*}
\caption{Detected sources at 610~MHz above the 7$\sigma$ level (first entries); the full table with 2796 records is available as online material.}
\centering
\begin{tabular}{l@{~~~}r@{~~~}r@{~~~}r@{~~~}r@{~~~}r@{~~~}r@{~~~}r} 
\hline\hline    
ID & $RA_{\rm J2000}$ & $Dec_{\rm J2000}$ & Total flux & Peak flux & $\theta_1$ & $\theta_2$ & $PA$ \\
   & (h,m,s) & (deg,$'$,$''$) & (mJy)  & (mJy)  &  ($''$)  & ($''$)  & (\degr) \\
\hline
BIC610-0001 & 20:11:32.02$\pm$0.009 & 40:53:10.84$\pm$0.10 & 14.5$\pm$0.69 &  7.8$\pm$0.26 & 9.3$\pm$0.34 & 9.3$\pm$0.34 &  63.3$\pm$3.83\\
BIC610-0002 & 20:11:33.03$\pm$0.001 & 40:53:18.92$\pm$0.02 & 63.8$\pm$0.75 & 42.0$\pm$0.24 & 7.9$\pm$0.06 & 7.9$\pm$0.06 &  60.6$\pm$1.35\\
BIC610-0003 & 20:11:34.94$\pm$0.009 & 41:31:49.27$\pm$0.09 & 12.5$\pm$0.63 &  7.9$\pm$0.26 & 8.8$\pm$0.33 & 8.8$\pm$0.33 & 112.7$\pm$2.90\\
BIC610-0004 & 20:11:48.15$\pm$0.019 & 40:51:32.52$\pm$0.25 &  2.3$\pm$0.40 &  2.3$\pm$0.23 & 6.6$\pm$0.73 & 6.6$\pm$0.73 &  53.2$\pm$23.77\\
BIC610-0005 & 20:11:53.58$\pm$0.007 & 40:48:39.49$\pm$0.11 &  8.6$\pm$0.50 &  6.4$\pm$0.24 & 7.4$\pm$0.29 & 7.4$\pm$0.29 & 128.0$\pm$11.67\\
BIC610-0006 & 20:11:55.45$\pm$0.008 & 42:13:40.96$\pm$0.11 &  4.3$\pm$0.33 &  4.1$\pm$0.19 & 6.3$\pm$0.30 & 6.3$\pm$0.30 & 106.9$\pm$27.75\\
BIC610-0007 & 20:11:56.63$\pm$0.014 & 42:13:37.89$\pm$0.24 &  2.4$\pm$0.34 &  2.2$\pm$0.19 & 6.7$\pm$0.60 & 6.7$\pm$0.60 & 142.6$\pm$23.84\\
BIC610-0008 & 20:11:58.89$\pm$0.009 & 41:47:49.24$\pm$0.18 &  2.0$\pm$0.27 &  2.6$\pm$0.18 & 5.6$\pm$0.42 & 5.6$\pm$0.42 &   1.1$\pm$21.54\\
BIC610-0009 & 20:11:59.38$\pm$0.007 & 40:28:25.33$\pm$0.09 &  7.6$\pm$0.44 &  6.7$\pm$0.23 & 6.8$\pm$0.25 & 6.8$\pm$0.25 &  57.4$\pm$11.20\\
BIC610-0010 & 20:12:00.79$\pm$0.018 & 42:02:34.40$\pm$0.28 &  2.1$\pm$0.34 &  1.8$\pm$0.17 & 6.7$\pm$0.69 & 6.7$\pm$0.69 & 134.7$\pm$70.97\\
\hline
\end{tabular}
\tablefoot{$\theta_1$, $\theta_2$, and $PA$ represent the elliptic source size and orientation, and correspond to the the major axis, the minor axis, and the position angle of the fit by the PyBDSM routines.}
\label{sources6}
\end{table*}


\section{Determining spectral indices} 

The spectral index $\alpha$ of a source is a key parameter in the determination of its nature. When the flux densities at frequency bands centered at $\nu_1$ and $\nu_2$ are $S_1$ and $S_2$, and $S_\nu \propto \nu^{\alpha}$, $\alpha = \log(S_1/S_2) / \log (\nu_1/\nu_2)$. For the observations we processed, $\nu_1 = 325$~MHz and $\nu_2 = 610$~MHz, and $\alpha$ can be derived when a source was detected at both bands. 

In the process of obtaining spectral index information, we verified whether the sources we detected at one of the observing bands were positionally coincident with one or more sources detected at the other band. 
To do this, we determined for each 325 MHz source ellipse whether it overlapped one or more of the 610 MHz source ellipses. When this was the case, we classified the overlapping as {\sl \textup{partial}} when the ellipse at 325 MHz partially overlapped the ellipse at 610~MHz, or as full when the 610 MHz ellipse was contained in the 325 MHz ellipse. For {partial} cases, we registered the percentage of overlapping area ($OA$). 

We then studied the 610 MHz ellipse/s that was/were related to each single 325 MHz ellipses and calculated a corresponding 610 MHz contributing flux $SC_{2}$ that we used in the spectral index expression in the following way. For { full} cases, we considered $SC_{2} = S_2$. For { partial} cases, we set $SC_{2} = S_2$ when $OA \geq 70\%$, $SC_{2} = 0.5\,S_2$ when $70\% > OA > 30\%$, and $SC_{2} = 0$ elsewhere. 
We found that 993 sources at 325 MHz overlap one or more 610 MHz sources, and we computed the corresponding spectral indices by considering for each source at 325~MHz all the overlapping sources at 610~MHz with the weights as explained above. Table\,\ref{tab:spixshort} lists the 325~MHz source  with its central coordinates, the 610~MHz source(s) that partially or fully overlap the former, and the spectral index $\alpha$ as derived from $S_1$ and $SC_2$ (a few entries; the full table is  available as online material). {We present the spectral index uncertainty by error propagation in the very conservative case, that is, using the flux density errors given by PyBDSF combined with a 10\% error for flux density scales (see Sect.\,\ref{datared}).}

To evaluate the probability of random matches when spectral indices are derived, we calculated the inverse of the number of sources per square degree over the area of the synthesized beam. At 325~MHz, we obtained that there will be one such a coincidence in 1700 sources. At 610~MHz, the probability of a random match is one source in 3230. We found 1048 sources at 325~MHz, and 2796 sources at 610~MHz, therefore we assume that is very unlikely that unrelated sources overlap in our sample.


\begin{table*}
\caption{Sources detected at both frequency bands (325 and 610~MHz) and spectral index information (first entries); the full table of  993 records is available as online material.}
\label{tab:spixshort}
\centering
\begin{tabular}{l c l r} 
\hline\hline    
ID at 325MHz & $RA, Dec_{\rm J2000}$ & ID at 610MHz & $\alpha_{\rm 610MHz}^{\rm 325MHz}$\\ 
             & (hms, dms)           &              & \\
\hline
BIC325-0002 & 20:13:56.45, 41:33:06.07 & BIC610-0104,-0105 &    -0.4$\pm$0.26\\
BIC325-0003 & 20:14:04.78, 40:56:14.47 & BIC610-0109 &    -0.7$\pm$0.32\\
BIC325-0004 & 20:14:07.82, 41:02:47.35 & BIC610-0112 &    -1.4$\pm$0.48\\
BIC325-0005 & 20:14:17.56, 41:29:24.16 & BIC610-0123 &    -1.1$\pm$0.23\\
BIC325-0006 & 20:14:18.62, 41:18:10.88 & BIC610-0124 &    -1.0$\pm$0.23\\
BIC325-0007 & 20:14:19.17, 41:29:36.95 & BIC610-0126 &    -0.9$\pm$0.23\\
BIC325-0008 & 20:14:24.35, 41:41:32.07 & BIC610-0133,-0134,-0131 &    -0.8$\pm$0.25\\
BIC325-0009 & 20:14:27.13, 41:17:52.59 & BIC610-0140 &    -1.2$\pm$0.34\\
BIC325-0010 & 20:14:29.42, 41:41:32.46 & BIC610-0146 &    -1.3$\pm$0.24\\
BIC325-0012 & 20:14:30.76 , 41:41:41.64 & BIC610-0147 &    -0.9$\pm$0.25\\
\hline
\end{tabular}
\end{table*}


\section{Catalog properties}

\subsection{Detections, flux densities, and noise levels}

   \begin{figure}
   \centering
  \includegraphics[width=9cm]{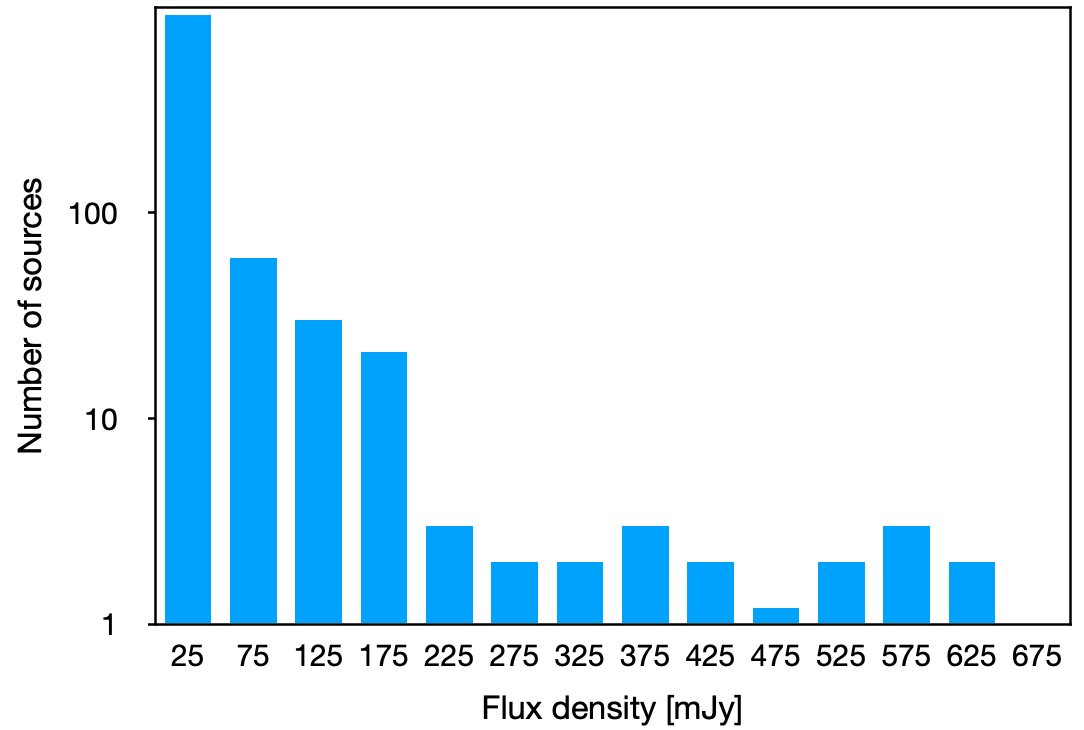} 
  \includegraphics[width=9cm]{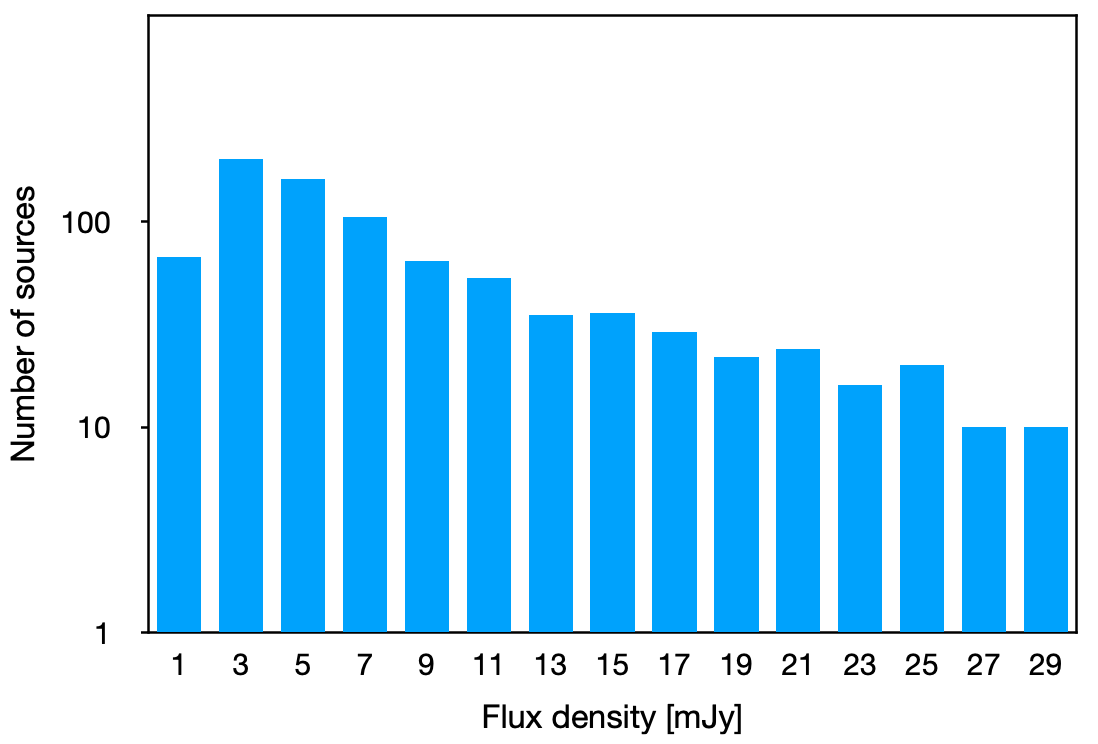}
      \caption{Top: Number of sources as a function of their integrated flux density  for 99.71\% of the sources cataloged at 325~MHz. Bottom: Same for flux densities up to 30~mJy (841 sources out of 1048, i.e., 80\%).}
         \label{fig:325-fluxhistos}
   \end{figure}

   \begin{figure}
   \centering
  \includegraphics[width=9cm]{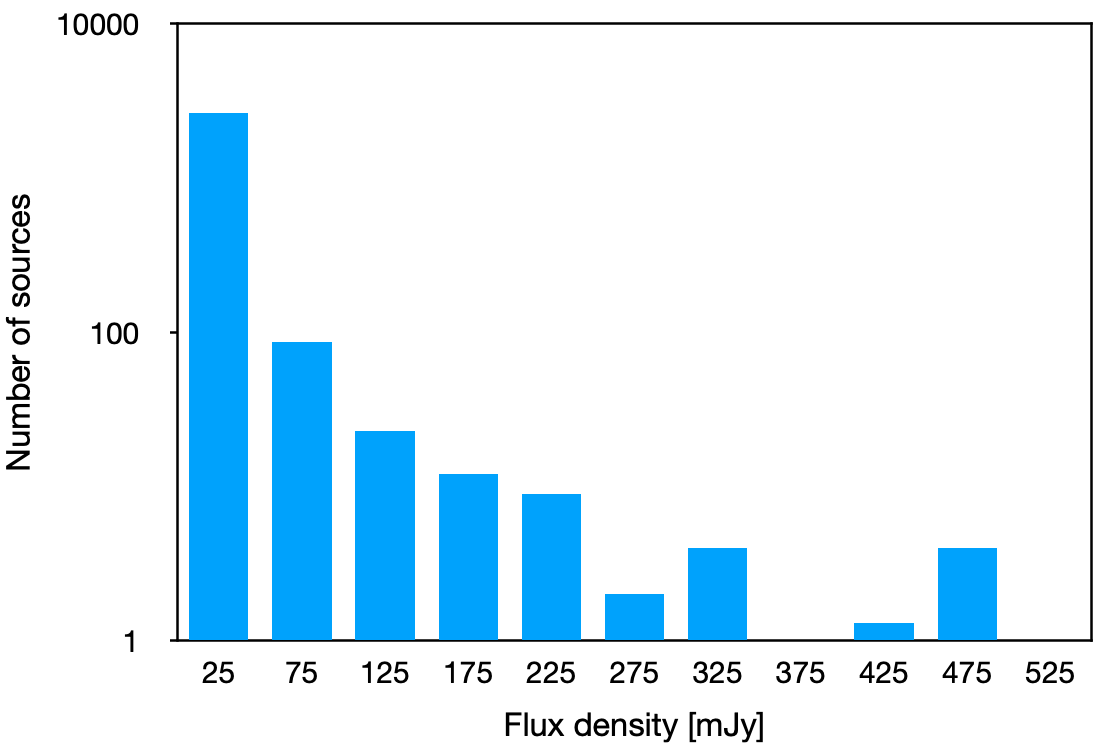} 
  \includegraphics[width=9cm]{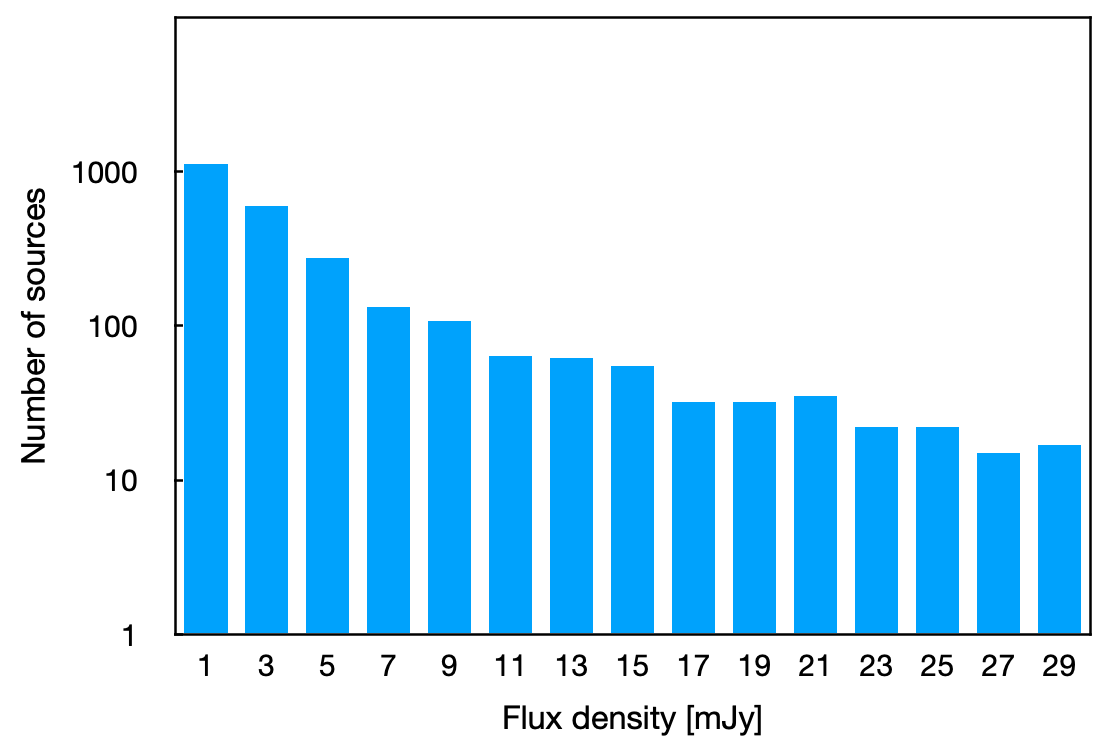}
      \caption{Top: Number of sources as a function of their integrated flux density for 99.75\% of the sources cataloged at 610~MHz. Bottom: Same for flux densities up to 30~mJy (2565 sources out of 2796, i.e., 91\%).}
         \label{fig:610-fluxhistos}
   \end{figure}

The catalog comprises 1048 sources at 325~MHz and 2796 sources at 610~MHz with flux densities greater than 7$\sigma$; here $\sigma$ represents the local rms noise at the source surroundings. The sources are characterized by their integrated and peak flux densities with corresponding errors, major and minor axes, and the position angle of a fitted ellipse, also with their errors. 

Figure\,\ref{fig:325-fluxhistos} displays the distribution of the flux density of the sources detected at the 325~MHz band. In the lower panel we present a zoom on the flux interval 0 -- 30~mJy, which contains 80\% of the sources. The corresponding histograms for 610~MHz appear in Fig.\,\ref{fig:610-fluxhistos}, with 91\% of the sources with fluxes up to 30~mJy. At both bands, the effect of favoring higher resolution (meanwhile outlining more discrete sources) by means of the weighting scheme is appreciable as a majority of sources with lower fluxes. The detail for the sources with lower flux shows the typical decrease with flux \citep[see, e.g., Fig.\,3 in][]{zoone1990}. The lower number of sources up to 5~mJy at 325~MHz is probably the effect of the detection threshold at this band ($\simeq$ 2~mJy on average).

We compared the number of sources we detected with the results from other studies. At the 325~MHz band, the survey by \citet{taylor1996}, carried out with the Westerbork Synthesis Radio Telescope (WSRT) at 327~MHz, reported 3984 sources over a detection threshold of 10~mJy, an area of 160 sq~deg, and an angular resolution $\geq 1'$, which means a ratio $R$ of 24.9 sources per square degree. At a similar frequency, we obtained 453 sources with fluxes above 10~mJy, thus a ratio of 40.1 sources per square~degree. The difference can be explained by the larger beam that was used by the former survey, which is six times larger than the synthesised beam we used because some nearby GMRT sources will be seen as one WSRT source. A quick comparison with the detections reported by \citet{setia2003} (synthezised beam larger than five times that of this research) found that more than 90\% of their sources that are present in the area are in common.

The VLA FIRST survey \citep{FIRSTsurvey}, performed at 20~cm, found 946432 sources above a detection threshold of 1~mJy (0.15~mJy rms) using angular resolution images of $\sim5"$ over an area of 10575~sq~deg, and then $R=89.5$. The FIRST detection limit corresponds to a value of 1.8~mJy when it is scaled to the 610~MHz band using a spectral index of $-0.7$. In our catalog, almost 1800 sources showed flux densities higher than 1.8~mJy and $R=91.4$, which agrees very well with the results from \citet{FIRSTsurvey}.

   \begin{figure}
   \centering
  \includegraphics[width=9cm]{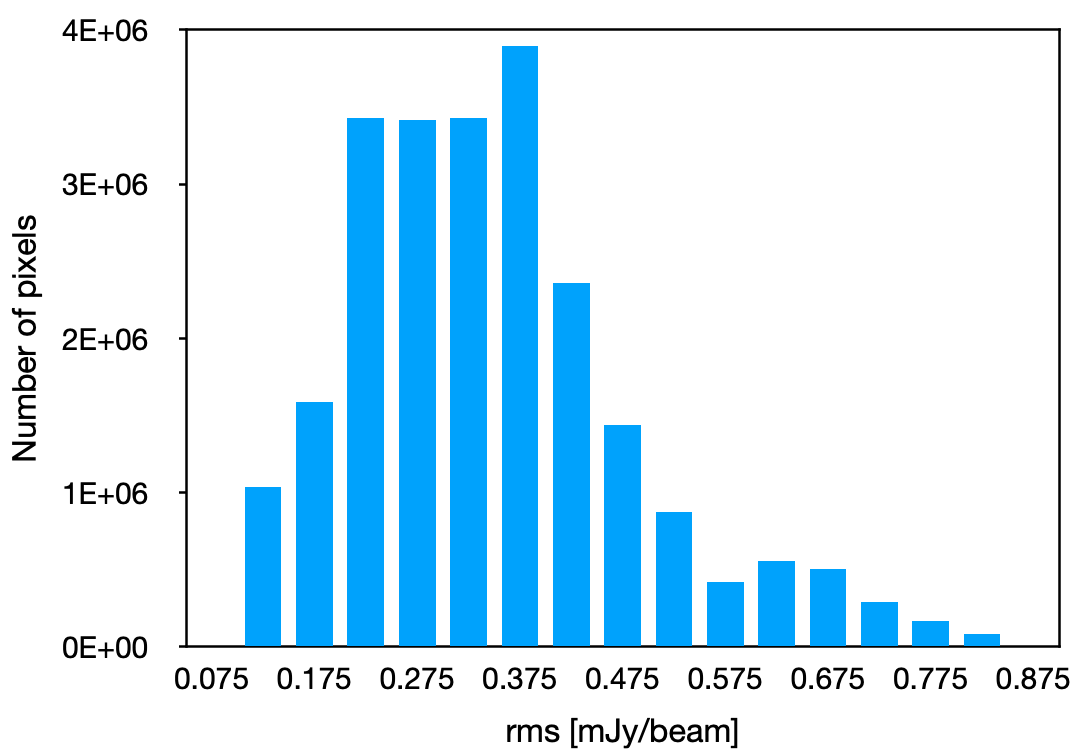}
  \includegraphics[width=9cm]{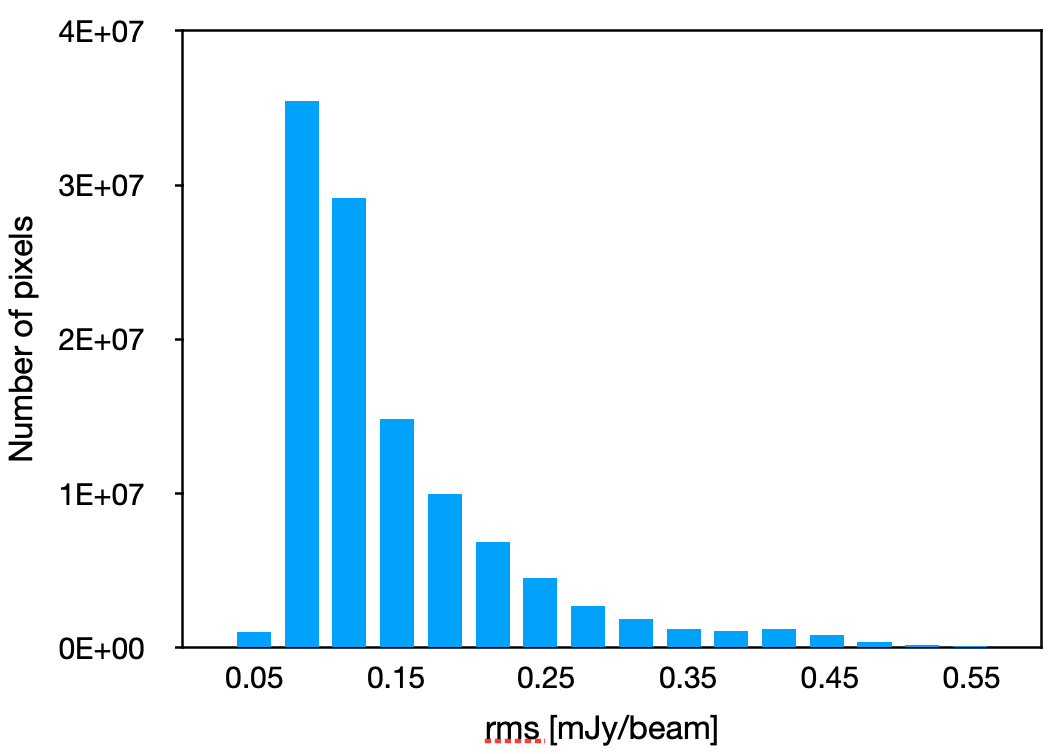}
      \caption{Distribution of the rms in the mosaics at 325~MHz (top panel) and at 610~MHz (bottom panel).}
         \label{fig:rmss-from-py}
   \end{figure}

   \begin{figure}
   \centering
  \includegraphics[width=9cm]{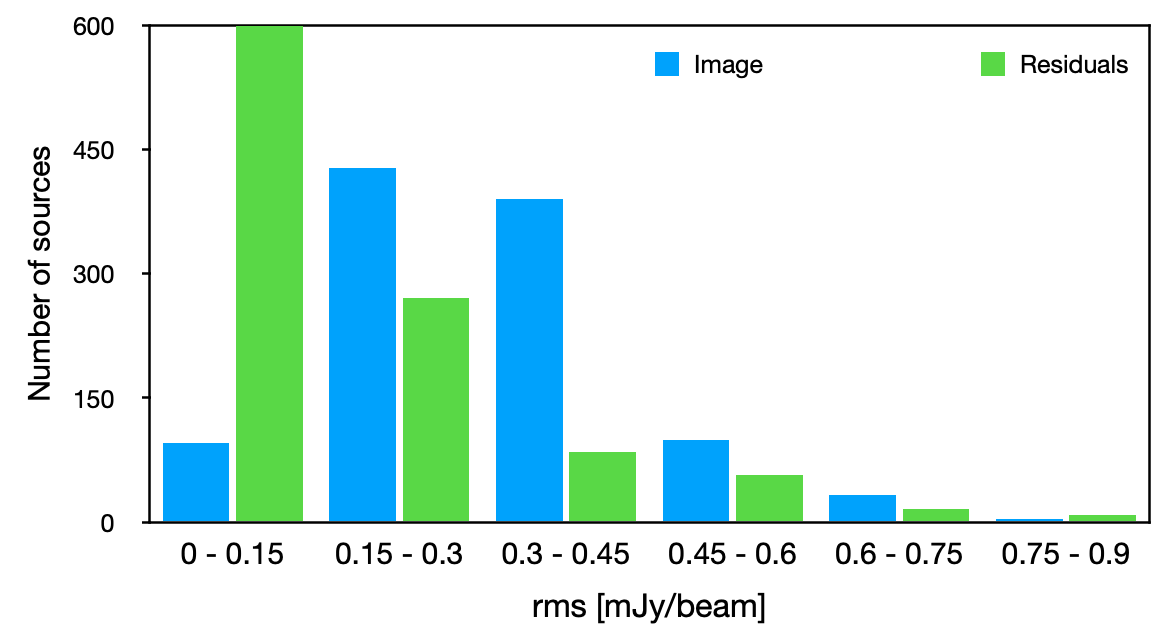}
  \includegraphics[width=9cm]{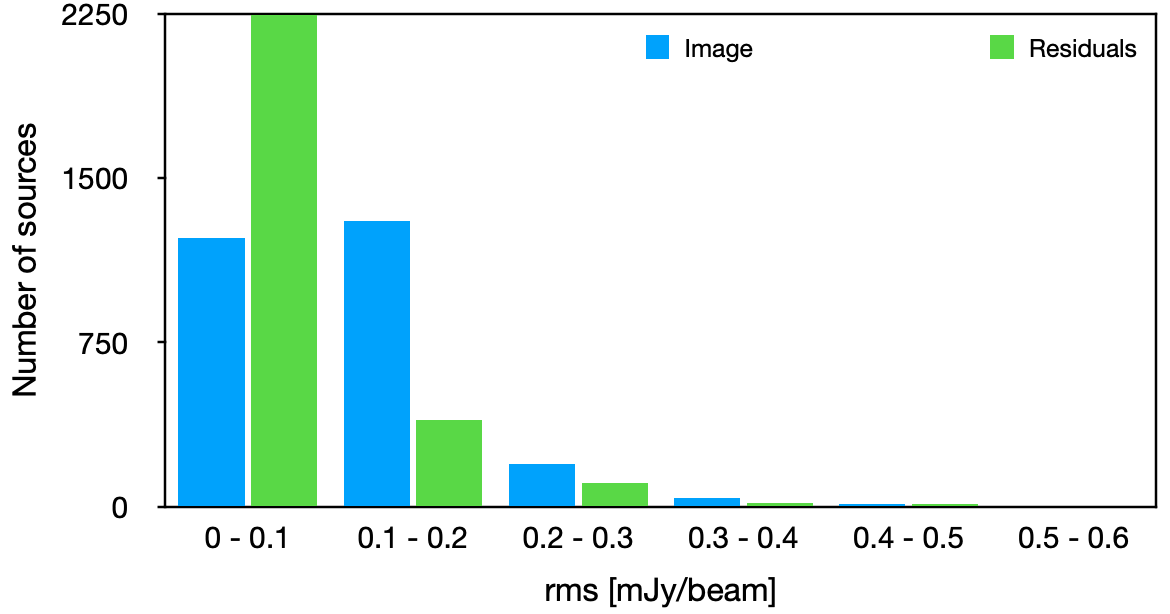}
      \caption{Distribution of the rms in the mosaics  before (light blue bars) and after source extraction (green bars) at 325~MHz (top panel) and at 610~MHz (bottom panel).}
         \label{fig:rms-comps}
   \end{figure}

For the rms per pixel at each band, we present in Fig.\,\ref{fig:rmss-from-py} the histograms showing the distribution of the rms in the fields we observed as obtained when the PyBDSF routines were applied at both bands. This rms value  is estimated by PyBDSF near each source before fitting. At 325~MHz, 88\% of the pixels show an rms of up to  0.5~mJy~beam$^{-1}$. At 610~MHz, 80\% of the pixels show an rms of up to  0.2~mJy~beam$^{-1}$. 
We also compared these rms values and the rms values of a residual image from which the  contribution of the  fitted sources was subtracted. The results are presented in Fig.\,\ref{fig:rms-comps}. The rms values decrease considerably, as expected.

\subsection{Resolved and unresolved sources}

\begin{figure}
   \centering
   \includegraphics[width=8cm]{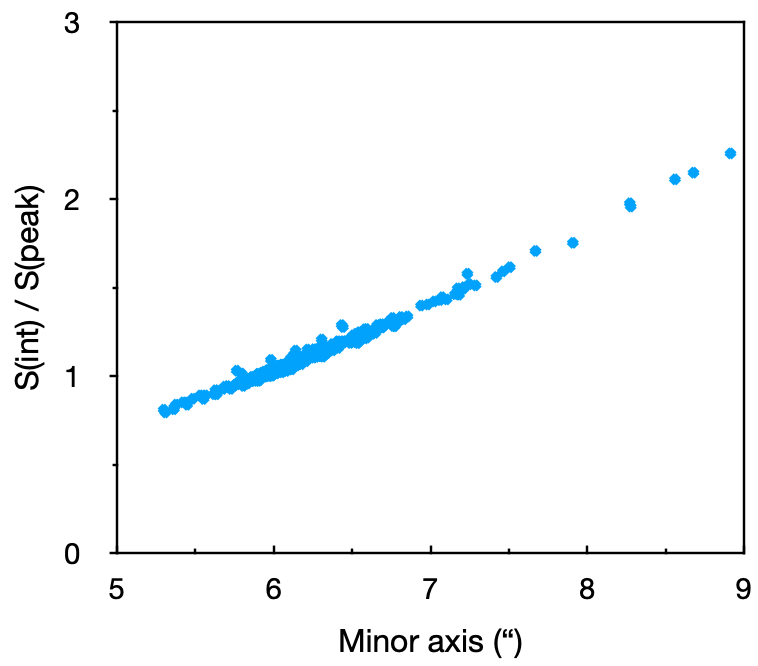}
   \includegraphics[width=8cm]{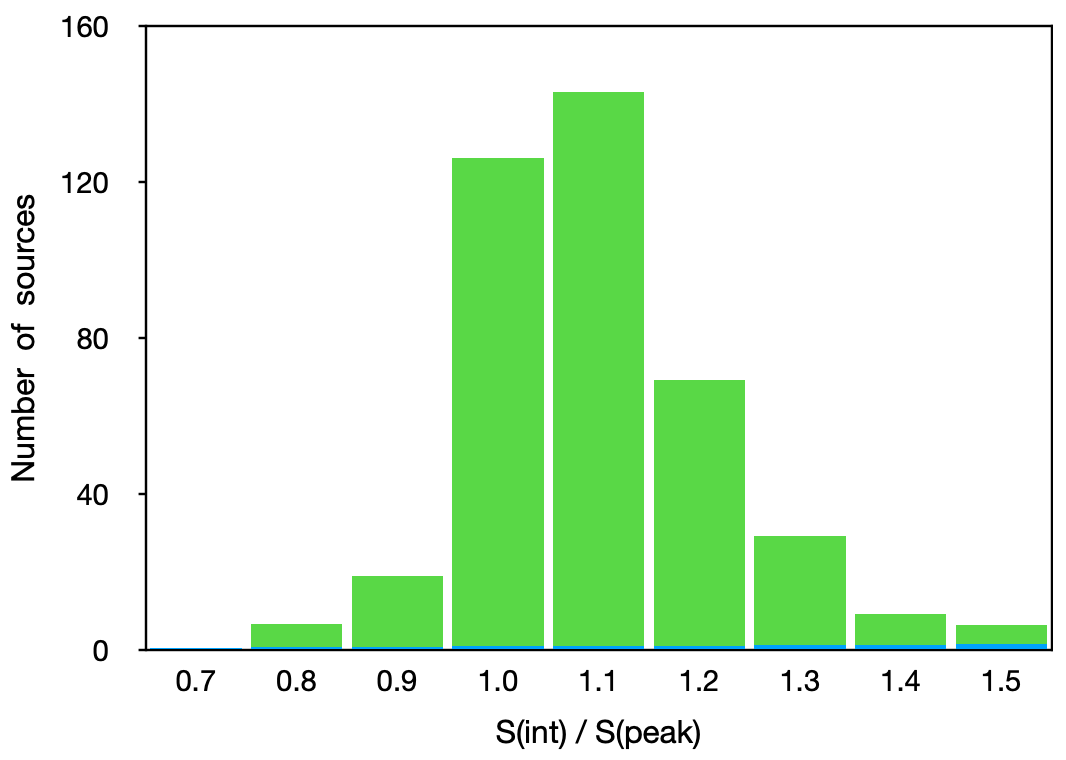}
      \caption{Top panel: Ratio of total flux over peak flux as a function of their minor axis for sources with $\theta_1 / \theta_2 \leq 1.05$, cataloged at 610~MHz. A power fit yields $S_{\rm int}/S_{\rm peak} = 1.25$ at $\theta_2 = 6.6"$. Bottom panel: Distribution of the flux ratio of the same group of sources.}
        \label{fig:unresolved610}
   \end{figure}

\begin{figure}
   \centering
   \includegraphics[width=9cm]{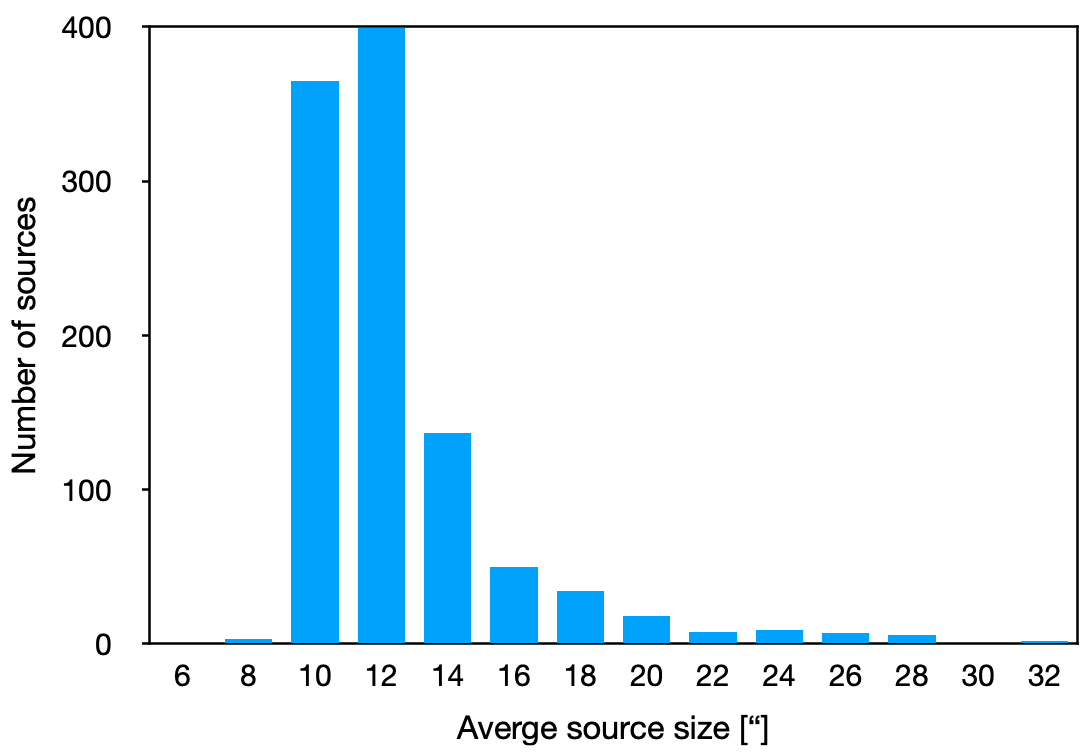}
      \caption{Distribution of the average axis [0.5\,$\times$\,(major axis + minor axis)] of the ellipses representing the sources at 325~MHz.}
         \label{fig:semiaxes325}
   \end{figure}

\begin{figure}
   \centering
   \includegraphics[width=9cm]{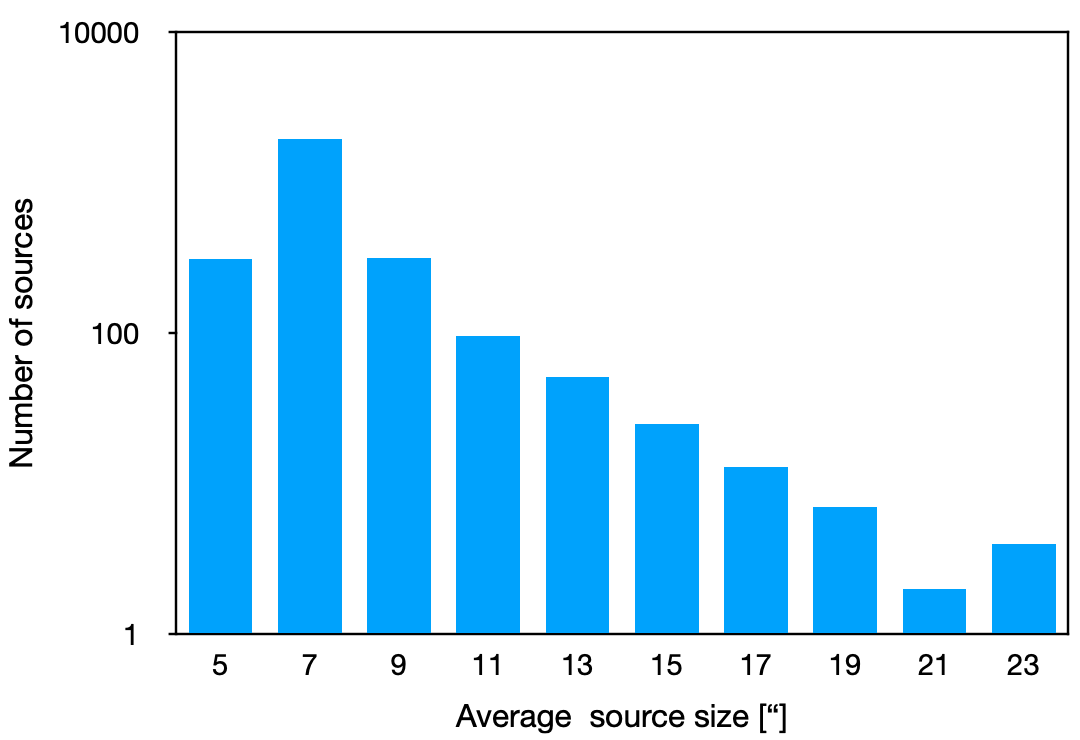}
      \caption{Distribution of the average axis [0.5\,$\times$\,(major axis + minor axis)] of the ellipses representing the sources at 610~MHz.}
         \label{fig:semiaxes610}
   \end{figure}

To distinguish between resolved and unresolved sources, we plot in Figure\,\ref{fig:unresolved610} the ratio of the total (integrated) flux over the peak flux for circular sources ($\theta_1 / \theta_2 < 1.05$, 418 sources) at 610~MHz. The ratio remains below 1.25 out to 6.6$"$, which we adopt as the dividing line between resolved and unresolved sources. This value by coincidence is the size of the synthesised beam plus a 10\% error at this frequency. 
We applied the same criterion based on the ratio value of 1.25 for the sources detected at 325~MHz.
The distribution of the mean axes (sizes) of the cataloged sources, in the form of the average of the major and minor axes of each ellipse, is shown in Figs.\,\ref{fig:semiaxes325} and \ref{fig:semiaxes610}.


\subsection{Source multiplicity}

\begin{figure}
   \centering
  \includegraphics[width=9cm]{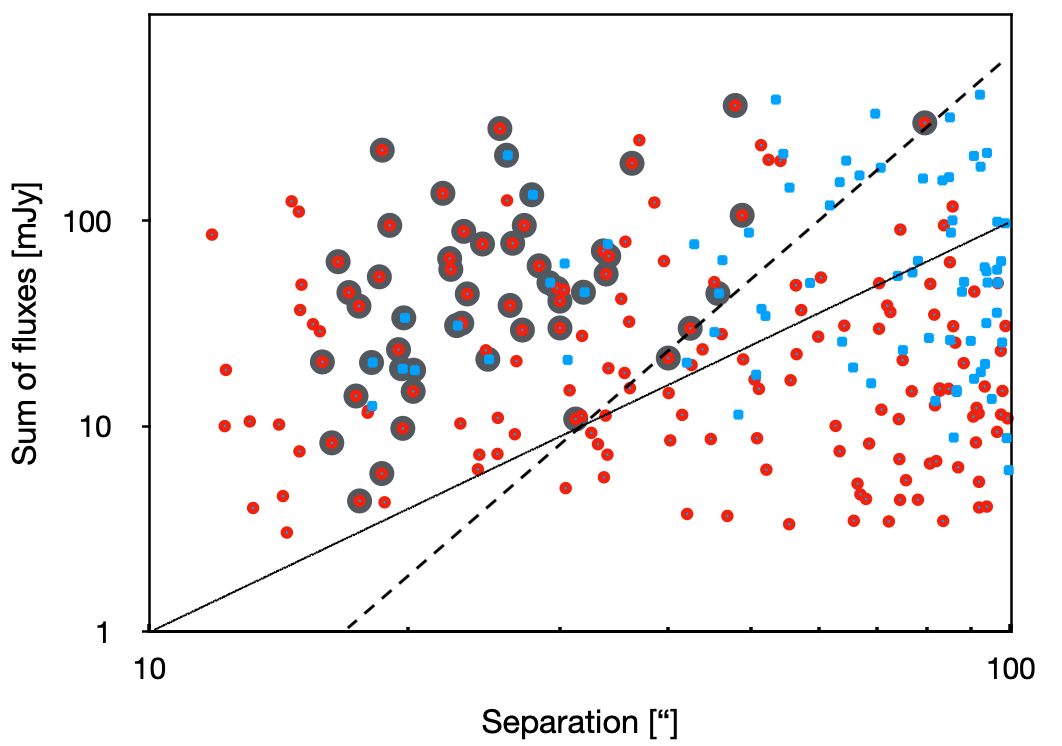}
  \includegraphics[width=9cm]{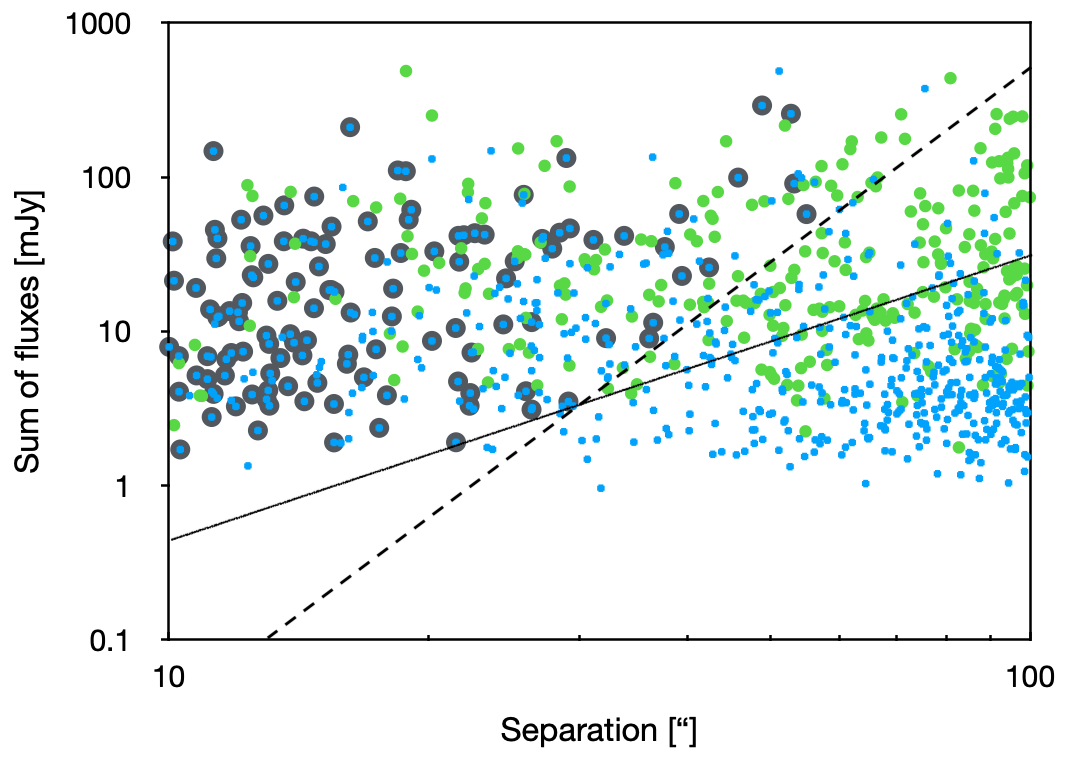}
      \caption{Sum of the flux densities from pairs of nearby sources ($FS$) vs. source separation ($x)$. Top panel: 325~MHz sources; in red, pairs of sources whose flux ratio (integrated/peak) is below 4; in blue, the rest; larger dark gray circles: pairs of confirmed double sources (see text). Dashed line: $(x/16)^{3.5}$. Thin solid line: $(x/10)^{2}$. Bottom panel: for 610~MHz sources; in green, pairs of sources whose flux ratio is below 4; in blue, the rest; larger dark gray circles: pairs of double sources (see text). Dashed line: $(x/20)^{3.5}$. Thin solid line: $(x/16)^{2}$.}
        \label{fig:multiplicities}
   \end{figure}

During the visual inspection process of all sources we found at both bands, we marked those that are characterized by adjacent emission components, fit by distinct Gaussian functions. In  many of them, even a bridge linking components was clearly seen.
Following the technique by \citet{maglio1998} and \citet{huynh2005}, we listed the sources that presented a companion up to 2$'$, and distinguished the pairs (source$+$companion) where the flux density ratio (brighter over weaker) remained below 4. Fig.\,\ref{fig:multiplicities} represents these groups in the plane of the sum of the fluxes ($FS$) of the components for each  pair versus the separation ($x$) between components.

Two areas are clearly visible in the plots at both bands, depending on whether they contain visually checked double pairs. Previous works have found that the limit between the areas can be described by $FS \propto x^{2}$.
The data presented here appear to  be better confined with an exponential of 3.5; see Fig.\,\ref{fig:multiplicities}, where we plot the two limiting lines. In principle, we can infer that the components of the pairs in the left areas are more likely to be physically related. 

\subsection{Considerations of the spectral indices}

\begin{figure}
   \centering
   \includegraphics[width=9cm]{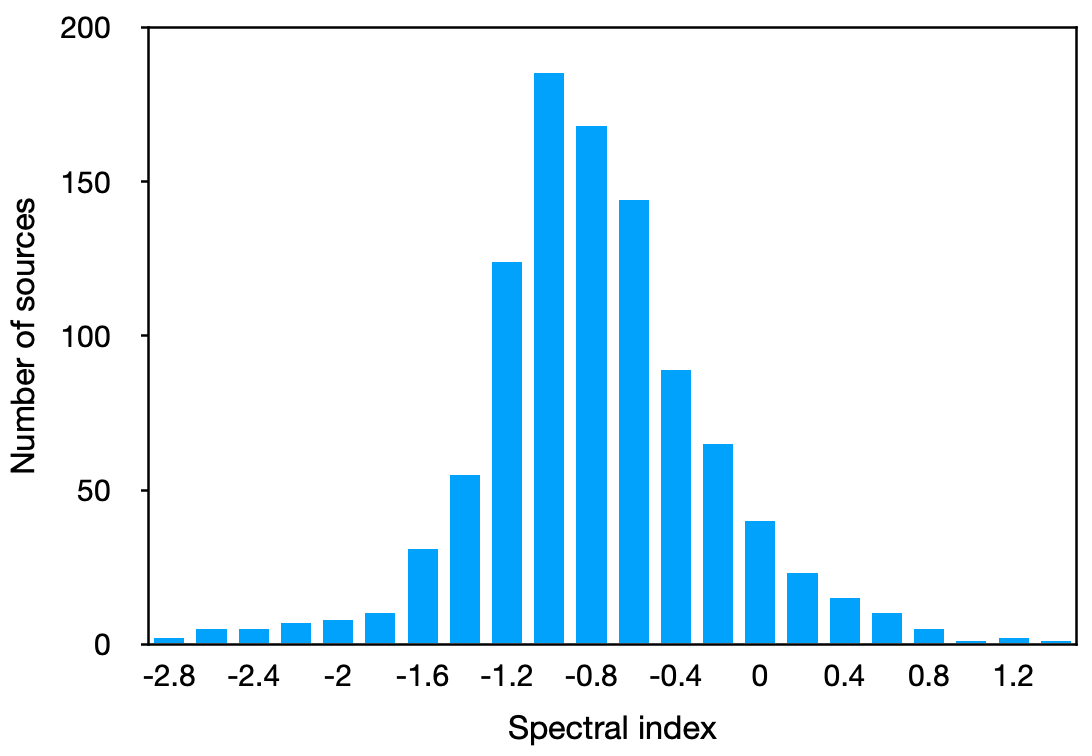}
      \caption{Distribution of spectral indices corresponding to the sources detected at both frequency bands {(see Table\,\ref{tab:spixshort})}.}
         \label{fig:spixes}
   \end{figure}

\begin{figure}
   \centering
   \includegraphics[width=9cm]{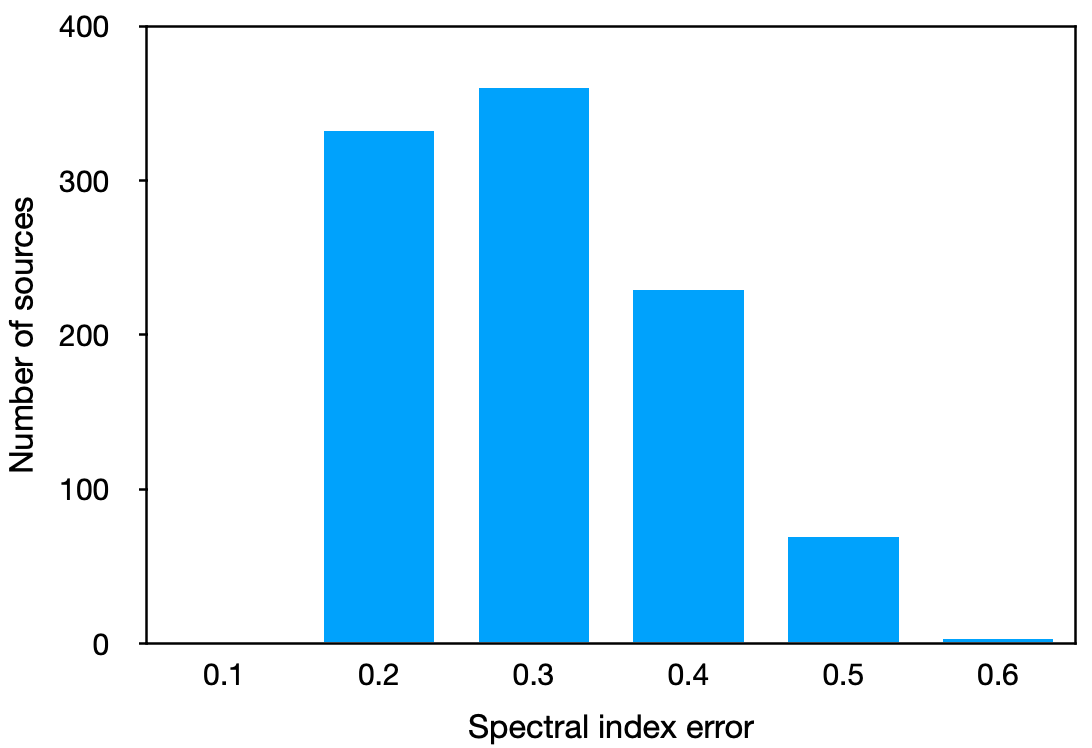}
      \caption{Distribution of spectral index errors corresponding to the sources detected at both frequency bands {(see Table\,\ref{tab:spixshort})}.}
         \label{fig:spixeserror}
   \end{figure}

In the case of the spectral index distribution of the $\text{about }$one thousand sources that are detected at two bands in our catalog, the pronounced maximum at $\alpha = -1$ confirms the nonthermal nature of most of the sources, see Figs.\,\ref{fig:spixes} and \ref{fig:spixeserror}.
The median error in $\alpha$ is {0.29}. The spectral index values span from $-3.06$ to $+1.41$. Only 4 out of 993 values are below $-2.5$. Variability can be one of the reasons for the extreme absolute values of the index: the two frequency data were taken at different times. They might also arise because we systematically considered the fluxes for sources that do not fully  overlap.

{The uncertainties on the spectral indices are somewhat large, which is due to the conservative error on the flux density and because the two frequencies lie only about a factor of two apart. However, this should be sufficient to broadly categorize the sources as  thermal, nonthermal, or pick up sources with very steep spectra.}

The 993 sources with a spectral index value detected at the 325~MHz band correspond to 1065 sources  at the 610~MHz band: at this latter frequency, the synthesized  beam is smaller, and we found two or even three 610~MHz sources that overlap  the same 325~MHz object; see\,Table\,\ref{tab:spixshort}.
The surveyed area at 325~MHz totaled 23486672 pixels with signal (pixel size = $2.5'' \times 2.5''$), or 11.3 sq deg. At 610~MHz, 113370869 pixels with signal accounted for 19.7 sq deg (pixel size = $1.5'' \times 1.5''$). In the same area covered by the 325 MHz mosaic, 1796 sources at 610~MHz (out of the total number of 2796) were fit, thus the ratio of source-fits at 610 to 325~MHz is 1.71. This can be explained by sensitivity limitations,  considering that at 610~MHz we have better angular resolution and lower noise, which allows us to detect thermal sources that will remain undetected at 325~MHz also because they are fainter, except for the cases when we picked up more than one source counterpart of a single 325~MHz source at 610 MHz.

\section{Search for counterparts}

After the catalog of 325 and 610~MHz sources was completed,
we searched for nearby objects as possible counterparts to the (1048$+$2796=) 3844 entries. We used the Simbad database\footnote{http://simbad.u-strasbg.fr/simbad/sim-fid}, with two input tables: one containing the coordinates of the records of the 325~MHz sources, and a second table with those of the 610~MHz sources. The search radius was set as the semimajor axis of the ellipse fit of each record. We found possible counterparts for 85 sources at 325~MHz and for 138 sources at 610 MHz, where more than one nearby object was found for some of the sources. We studied all possible counterparts in each case and also verified that the nearby objects were inside the fitted ellipse, that is, taking the semiminor axis and position angle of the source fit into consideration. We gathered the findings in Table\,\ref{counterparts}, which presents possible counterparts for 5 sources that are only detected at 325~MHz, for 52 sources that are only detected at 610~MHz, and for 86 sources that are detected at both bands, ordered by right ascension. The angular distance $d$ between the GMRT source and the potential counterpart and the spectral index, if applicable, is also listed. 
By searching the literature, we investigated the nature of the potential counterpart, and propose the more plausible object that could be associated with the GMRT sources reported here whenever possible, along with its reference or, in the worst case, the reference of flux measured at other wavelength(s). In the cases with preexisting 325 MHz observations, we quote no counterpart because our observations superseded them in sensitivity or also in angular resolution. In addition to information from the surveys mentioned in Sect.\,\ref{sec:cyg}, valuable material was found in \cite{vollmer2010}, who compiled flux values of sources at the radio range including those of the Cygnus region relevant here. An exception is made for fluxes at the 610~MHz band, for which no previous data were found. In this sense, our catalog completes many radio spectra and provides 610 MHz flux values of $\sim$2800 sources for the first time.

Of the sources with possible counterparts that were detected at both frequencies, 16\% have flat or positive spectral index (nominally, $\alpha > -0.2$). Their counterparts are mostly either stars, young stellar objects, or radio sources detected at higher frequencies.

At the 325~MHz band, 8\% of all the 325~MHz cataloged sources have a possible counterpart. At 610~MHz,  5\% of all the 610~MHz cataloged sources have a possible counterpart. These low percentages can be explained considering that at the observed bands, decimeter wavelengths, we mostly sample nonthermal sources, at which high-energy (HE) processes presumably take place. These might be potential counterparts to the HE sources; but the angular resolutions and sensitivities for instruments working at HE ranges are higher and lower, respectively, precluding successful cross-identifications between radio and HE sources. 

{Preliminary results of a study focused on the unresolved unidentified sources with negative spectral index ($\sim$340 sources of the present catalog), such as was performed by \citet{chakra2020}, indicated that differential source counts in the distribution trend of extragalactic sources result from other surveys or catalogs. 
However, we recall that the observed fields are in the surroundings of the Galactic plane, dense in Galactic sources, both of thermal and nonthermal nature. 
A detailed investigation to distinguish in which proportion, and more interesting, from which types of objects, Galactic and extragalactic sources contribute etc. is beyond the scope of this paper.
}

\section{Related studies and prospects}

The GMRT observations that gave rise to our catalog allowed us simultaneously to carry out research on individual populations of astronomical objects. Specifically, different types of objects that can produce nonthermal emission were or are being studied in separated investigations: AGNs and two-lobed sources, counterparts to HE sources, massive early-type stars \citep{cygnusstars2020}, protoplanetary disk-like sources \citep{isequilla2019} and young stellar objects 
\citep{isequillaBAAA}.
Finally, the survey images will be presented elsewhere. Future work includes the scrutiny of sources between 3 and 7$\sigma$. The corresponding source extraction, after a thorough validation process, of such a large area with the angular resolution of a few arcseconds provided by the GMRT data at decimeter frequencies will certainly reveal a plethora of interesting objects and  powerful statistical results of the nonthermal sky.

\begin{acknowledgements}
The authors are grateful to the referee, whose comments and
suggestions resulted in improving the analysis and presentation
of the article. The GMRT is operated by the National Centre for Radio Astrophysics of the Tata Institute of Fundamental Research. We thank the staff of the GMRT that made these observations possible. PB acknowledges support from ANPCyT PICT 0773--2017, and the contacts at NCRA, Pune for a very pleasant stay.
 ICCH acknowledges the support of the Department of Atomic Energy, Government of India, under the project  12-R\&D-TFR-5.02-0700.
 This research has made use of the SIMBAD database, operated at CDS, Strasbourg, France, and of NASA's Astrophysics Data System bibliographic services. 
\end{acknowledgements}

\bibliographystyle{aa} 
\bibliography{catalog-cyg} 

\clearpage
\onecolumn


\begin{landscape}
\begin{longtable}{l@{~~~}l@{~~~}r@{~~~}l@{~~~}l@{~~~}r@{~~~}l@{~~~}l@{~~~}l}
\caption{Counterparts of detected sources at 325~MHz and 610~MHz.}\\
\hline\hline
325-ID & 610-ID & Spectral &$RA_{\rm J2000}$ & $Dec_{\rm J2000}$ & $d$ & SIMBAD nearest &  Reference & Information on \\
BIC325-    &    BIC610-    & index & (h,m,s) & (deg,$'$,$''$) & ($''$)& source  &  & possible nature \\
\hline
\endfirsthead
\caption{continued.}\\
\hline\hline
325-ID & 610-ID & Spectral& $RA_{\rm J2000}$ & $Dec_{\rm J2000}$ & $d$ & SIMBAD nearest &  Reference & Information on \\
BIC325-    &    BIC610- & index   & (h,m,s) &  (deg,$'$,$''$) & ($''$)& source & & possible nature \\
\hline
\endhead
\hline
\endfoot
 ----- & 0045 & ----- & 20:12:57.27 & 41:51:49.34 & 4.6 & 2MFGC\,15386 & \cite{mitronova2004} & disk-like galaxy\\
  ----- & 0093 &-----  & 20:13:47.77 & 41:10:02.13 & 7.6 &  CXO\,J201348.3$+$411007.1 & \cite{montes2015} & X-ray source \\ 
0006 & 0124 & $-1.0\pm0.1$ &20:14:18.68 & 41:18:11.26 & 5.4 & CXO\,J201419.1$+$411813.4 & \cite{montes2015} & star-forming region \\
  ----- & 0217 & ----- &  20:15:21.60 & 40:34:43.93 & 6.8 & PN\,KjPn\,2 & \cite{kohoutek2001} & planetary nebula \\ 
0107  & 0429 & $-1.0\pm0.3$ & 20:17:54.12 & 41:24:14.49 & 5.3 & J201754.59$+$412413.98 & \cite{kryukova2014} & IR source\\
0119 & 0462 & $-1.6\pm0.2$ & 20:18:07.79 & 41:10:41.13 & 5.2 &  FGL\,J2018.1$+$4111 & \cite{abeysekara2018} & $\gamma$-ray source\\
0134 & 0531 & $-2.4\pm0.1$ & 20:18:38.18 & 40:41:00.42 & 3.7 & MSX6C\,G078.0875$+$02.6408 & \cite{panessa2015} & Seyfert galaxy\\
 ----- & 0543 &-----  &  20:18:42.69 & 44:17:34.13 & 2.2 &  NVSS\,J201842$+$441736 & \cite{vollmer2010} & \\
 ----- & 0551 & ----- & 20:18:46.65 & 42:20:04.91 & 8.2 & NVSS\,J201847$+$422010 & \cite{NVSS} & \\  
  ----- & 0582 & ----- & 20:19:02.55 & 40:18:26.02 & 4.1 & WSRTGP\,2017$+$4009 & \cite{vollmer2010} &\\
0151 & 0624 & $-0.2\pm0.3$ &20:19:19.30 & 40:54:51.52 & 7.8 & MTK2011--F1\,14  & \cite{melikian2011} & emission-line star\\
0162/0163/0164 & 0665/0666/0672 & $-1.0\pm0.2$ & 20:19:36.50 & 40:58:50.00 & $\sim$10 & NGR2010--VLAN\,2,3 & \cite{neria2010} & UC HII region \\
0165 & 0676 & $+0.9\pm0.1$ &20:19:38.89 & 40:56:36.21 & 1.3 & IRAS\,20178$+$4046--VLAN\,4  & \cite{neria2010} & UC HII region\\
0170 & 0690 & $-1.0\pm0.1$ &20:19:49.33 & 42:00:12.26 & 5.0 & J201949.77$+$420011.20 & \cite{kryukova2014} & YSO \\
 ----- & 0731 &----- & 20:20:09.10 & 43:40:22.78 & 2.0 & RX\,J2020.0$+$4357  & \cite{brinkmann1997} & X-ray source\\  
0182 & 0757 & $+0.1\pm0.2$ & 20:20:18.62 & 40:58:03.26 & 5.3 & J202019.08$+$405802.18 & \cite{kryukova2014} & YSO\\
 -----  & 0787 & ----- & 20:20:27.95 & 43:51:13.88 & 2.4 &  WR\,140 & \cite{pacwbcat} & WR system \\ %
0196 & 0808 & $+0.3\pm0.1$ & 20:20:35.65 & 40:57:54.84 & 6.2 & J202036.15$+$405753.58 & \cite{kryukova2014} & YSO\\
 $\,\,\,\,''$ & $\,\,\,\,''$ & $''$ & 20:20:36.18 & 40:57:53.08 & 6.6 & NVSS\,J202036$+$405754 & \cite{vollmer2010} & \\
 ----- & 0825 & ----- & 20:20:42.59 & 42:16:46.54 & 8.0 & MITG\,J2020$+$4216 & \cite{vollmer2010} & \\
0200 & 0842 & $-0.0\pm0.3$ &20:20:51.07 & 41:22:05.96 & 5.6 & J202051.55$+$412204.76 & \cite{kryukova2014} & YSO\\
 ----- & 0861 &-----  & 20:20:57.67 & 44:41:29.82 & 0.9 & NVSS\,J202057+444130 & \cite{vollmer2010} & \\
  ----- & 0894 &-----  & 20:21:18.24 & 41:19:59.65 & 5.0 & J202118.68$+$411958.86 & \cite{kryukova2014} & YSO \\
 -----  & 0967 &-----  & 20:21:49.02 & 44:00:37.37 & 2.7 & 2MASX\,J20214907$+$4400399 & \cite{veroncetty2010} & Seyfert galaxy\\
 ----- & 1049 &-----  & 20:22:17.22 & 42:24:49.47 & 10.4 & 18P\,22  & \cite{vollmer2010} & \\
0263 & 1051 & $-0.1\pm0.2$ & 20:22:17.85 & 43:53:01.09 & 8.3 & 2MASS\,J20221736$+$4353074 & \cite{pricewhelan2018} & red giant star\\
 ----- & 1112 &----- & 20:22:44.55 & 41:45:17.41 & 5.9 & J202245.07$+$414517.98 & \cite{kryukova2014} & YSO \\
 ----- & 1117 &----- & 20:22:46.33 & 41:07:00.44 & 5.8 & J202245.07$+$414517.98 &  \cite{kryukova2014} & YSO\\
 ----- & 1126 &----- & 20:22:52.35 & 44:48:20.56 & 3.3 &   BD$+$44\,3444 & \cite{elyajouri2016} & B8 star\\
 ----- & 1205 &----- & 20:23:19.06 & 43:12:44.08 & 8.2 & IRAS\,20216+4303 & \cite{vollmer2010} & \\
0343 & 1248 & $+0.1\pm0.1$ & 20:23:35.68 & 41:25:26.43 & 3.3 & J202335.81$+$412523.53 & \cite{kryukova2014} & YSO \\
0347 & 1253 & $-0.8\pm0.1$ & 20:23:39.05 & 44:01:04.49 & 10.6 & 18P\,25 & \cite{vollmer2010} & \\
0393 & 1329  & $-1.2\pm0.2$ &20:24:11.53 & 41:43:24.47 & 9.1 & TYC\,3160--519--1 & \cite{tychocat2000} & star\\
0390/0395 & 1334 & $-0.5\pm0.1$ & 20:24:15.43 & 43:22:32.22 & 5 & NVSS\,J202415$+$432235 & \cite{vollmer2010} &\\
0434 & 1404 & $-0.6\pm0.3$ & 20:24:46.19 & 42:23:13.16 & 9.3 & BD$+$41\,3737 & \cite{paunzen2015} & star \\
 -----  & 1420 &-----  &20:24:52.00 & 40:40:25.21 & 5.6 & G078.779$+$01.693 & \cite{anderson2015} & HII region\\
0460 & 1447 & $-0.7\pm0.1$ &20:25:00.59 & 41:48:25.43 & 6.1 & NVSS\,J202501$+$414829 & \cite{NVSS} & \\
0490 & 1493 & $+0.6\pm0.1$ & 20:25:19.02 & 43:35:19.40 & 4.7 &  J2025$+$4335 & \cite{immer2011} & \\  
0496 & 1503 & $-0.8\pm0.1$ & 20:25:22.74 & 44:19:33.44 & 1.4 & NVSS J202522$+$441934  & \cite{NVSS} & \\
0503 & 1511 & $-0.8\pm0.2$ & 20:25:24.78 & 41:03:19.48 & 3.6 & G079.151+01.830 & \cite{solin2012} & star-forming region \\ 
0526 & 1557/1563 & $-1.1\pm0.1$ & 20:25:40.53 & 42:32:17.14 & 6.4 & NVSS\,J202540$+$423222 & \cite{NVSS} & \\
 ----- & 1569 &----- & 20:25:42.91 & 41:56:15.55 & 5.6 & NVSS\,J202543$+$415618  & \cite{vollmer2010} &\\
 ----- & 1572 &----- & 20:25:44.10 & 41:56:02.14 & 6.0 & J202544.53$+$415605.70 & \cite{kryukova2014} & YSO \\
0600 & 1680 & $-1.0\pm0.1$ & 20:26:25.48 & 42:32:09.42 & 5.2 & NVSS\,J202625$+$423214 & \cite{NVSS} &\\
0604 & 1685 & $-0.8\pm0.1$ &20:26:26.26 & 44:39:27.16 & 3.9 &  NVSS\,J202625$+$443927 & \cite{vollmer2010} & \\
0626 & 1728 & $-0.9\pm0.1$ & 20:26:42.98 & 40:51:28.03 & 10.5 & NVSS J202642+405138 & \cite{vollmer2010} & \\
 -----  & 1821 &----- & 20:27:18.99 & 40:25:00.97 & 1.0 & WSRTGP\,2025+4015 & \cite{vollmer2010} &\\
0684 & 1822  & $-1.1\pm0.1$ & 20:27:19.57 & 43:13:58.44 & 14.3 & TYC 3164--341--1 & \cite{tychocat2000} & star\\
0722 & 1908 & $-0.8\pm0.1$ & 20:27:52.92 & 41:35:05.03 & 5.2 & J202753.37$+$413506.03 & \cite{kryukova2014} & IR source\\
0735 & 1944 & $-0.4\pm0.1$ & 20:28:04.06 & 41:13:54.24 & 5.1 & J202804.51$+$411354.56 & \cite{kryukova2014} & YSO\\
0744 & 1957 & $-0.8\pm0.1$ & 20:28:07.20 & 41:13:50.64 & 7.9 & UVEX\,J202807.55$+$411357.7 & \cite{verbeek2012} &white dwarf\\
 ----- & 1967 &----- & 28:28:10.38 & 45:12:51.73 & 0.2 &  NVSS\,J202810$+$451251 & \cite{NVSS} &\\
 ----- & 1980 &----- & 20:28:13.29 & 40:16:52.83 & 3.6 & WSRTGP\,2026$+$4006 & \cite{vollmer2010} &\\
0769 & 2013 & $-1.1\pm0.1$ &20:28:24.34 & 40:37:50.03 & 3.9 & 19P\,6 & \cite{wendker1991} &\\
 ----- & 2027 &----- & 20:28:31.01 & 40:59:59.05 & 3.9 & J202831.35$+$405959.24 & \cite{kryukova2014} & YSO\\
0788 & 2068 & $-1.1\pm0.1$ &20:28:49.60 & 41:18:37.36 & 0.6 & TYC\,3160--1079--1 & \cite{tychocat2000} & star\\
0789 & 2070 & $-1.3\pm0.1$ &20:28:50.79 & 41:34:37.13 & 5.5 & NVSS\,J202851$+$413438 &  \cite{vollmer2010} & \\
0804 & 2108 & $-1.1\pm0.1$ &20:29:00.90 & 42:12:59.28 & 5.5 & NVSS\,J202901$+$421259 & \cite{NVSS} & \\
0806 & 2115 & $-0.9\pm0.1$ &20:29:04.20 & 41:00:05.78 & 4.1 & 19P\,7  & \cite{wendker1991} & \\
0807 & 2118 & $-1.0\pm0.1$ &20:29:04.70 & 42:35:27.90 & 4.4 & TYC\,3160--1447--1 & \cite{tychocat2000} & star\\
-----  & 2159 &----- & 20:29:23.53 & 40:11:09.57 & 4 & AFGL\,2591 & \cite{jhonston2013} & star-forming region \\  
 ----- & 2170 &----- & 20:29:28.60 & 40:57:24.41 & 7.0 & 19P\,9 & \cite{wendker1991} & \\ 
 ----- & 2174 &----- & 20:29:30.94 & 41:34:20.72 & 5.2 & J202931.39$+$413421.96 & \cite{kryukova2014} & YSO\\
-----  & 2189 &----- & 20:29:36.50 & 41:51:40.89 & 4.6 & WSRTGP\,2027$+$4141 & \cite{vollmer2010} &\\
0846 & 2215 & $-1.7\pm0.3$ & 20:29:49.42 & 41:43:13.66 & 16.7 & ** GRV 344 & \cite{greaves2004} & stellar system\\
 ----- & 2223 &----- & 20:29:52.06 & 40:48:45.82 & 2.7 & G079.4430$+$01.0047 & \cite{urquhart2009} & HII region \\  
0854 & --- & ----- & 20:29:59.18 & 41:16:44.20 & 7.0 & IRAS\,20286$+$4105 & \cite{ramachandran2017} & star-forming region\\
0867 & 2280  &$-0.9\pm0.1$ & 20:30:14.17 & 40:41:08.67 & 10.4 & IRAS\,20283$+$4031 & \cite{parthasarathy1992} & star\\
0872 & 2299 &$-1.2\pm0.1$ & 20:30:30.47 & 44:12:26.40 & 4.9 & 2MASS\,J20303018$+$4412301 & \cite{pricewhelan2018} & star\\
0873 & 2311 & $-0.7\pm0.1$ & 20:30:36.58 & 41:06:06.10 & 6.2 & NVSS\,J203032$+$410634 & \cite{williams2013} & \\ 
 ----- & 2313 &----- & 20:30:37.68 & 42:20:57.96 & 15.9 & WSRTGP\,2028$+$4211 & \cite{vollmer2010} & \\
0874 & 2319 & $-0.8\pm0.1$ &20:30:39.72 & 41:23:31.73 & 0.4 & NVSS\,J203039$+$412331 & \cite{vollmer2010} &\\
-----  & 2351 &----- & 20:30:57.60 & 43:08:05.33 & 6.9 & WSRTGP\,2029$+$4257 & \cite{vollmer2010} &\\
0882 & 2370 & $-2.1\pm0.3$ &20:31:10.22 & 40:58:53.80 & 2.5 & J203110.44$+$405853.94 & \cite{kryukova2014} & YSO\\
 ----- & 2376 &----- & 20:31:14.29 & 42:22:42.95 & 4.1 & TYC\,3161--82--1 & \cite{tychocat2000} & star\\
0889 & 2381 & $-1.0\pm0.2$ &20:31:18.69 & 41:09:25.16 & 10.7 & RLP 933 & \cite{reddish1966} & star\\
 ----- & 2382 &----- & 20:31:19.14 & 40:18:09.90  & 0.4 & IRAS\,20293$+$4007\,VLA\,3 & \cite{sanchezmonge2008} & \\  
0890 & 2384 & $-1.0\pm0.1$ &20:31:19.90 & 40:40:56.04 & 1.6 & GPSR\,079.501$+$0.704  & \cite{zoone1990} & \\
 ----- & 2406 &----- & 20:31:37.33 & 40:22:58.79 & 0.3 & G79.29$+$0.46 & \cite{higgs1994} & wind shell\\  
 ----- & 2409 &----- & 20:31:39.72 & 40:16:08.36 & 0.2 & G79.29$+$0.46 & \cite{higgs1994} & wind shell \\
 ----- & 2420 &----- & 20:31:51.59 & 41:31:18.60 & 2.8 & BDB2006--234  & \cite{paredes2008} & X-ray star\\
0906 & 2432 & $-1.4\pm0.1$ &20:32:00.50 & 41:36:58.45 & 1.0 & J203201.7$+$413722 & \cite{paredes2008} & galaxy S lobe\\ 
0908 & 2435 & $-1.4\pm0.1$ &20:32:01.91 & 41:37:47.43 & 1.0 & J203201.7$+$413722 & \cite{paredes2008} & galaxy N lobe\\
0915 & 2448 & $-2.1\pm0.4$ &20:32:12.92 & 41:27:24.01 & 2.3 & MT91--213 & \cite{chen-am2019} & Be star+pulsar \\
 ----- & 2452 &----- & 20:32:14.13 & 40:42:24.88 & 0.7 & NVSS\,J203214$+$404226 & \cite{vollmer2010} & \\
 ----- & 2463 &----- & 20:32:21.17 & 40:17:18.63 & 0.5 & DR\,15\,I & \cite{colley1980} & part of nebula\\
0921 & 2464 & $-0.5\pm0.2$ &20:32:22.32 & 41:18:19.28 & 1.2 & Cyg\,OB2\,5 & \cite{pacwbcat} & OB stellar system \\
0924 & 2469 & $+1.4\pm0.1$ &20:32:25.65 & 40:57:28.09 & 1.5 & WR\,145a & \cite{agile2nd2019} & HMXB \\
0925 & 2470 & $-0.5\pm0.1$ &20:32:26.77 & 41:04:33.28 & 1.3 & HSC\,N & \cite{josep2006} & part of cloud\\  
0927 & 2472 & $-1.3\pm0.2$ &20:32:29.25 & 41:35:07.36 & 1.7 & 2MASS\,J20322935$+$4135061 & \cite{2MASScat} & IR source\\
0928 & 2473 & $-0.1\pm0.3$ &20:32:29.52 & 40:38:49.65 & 0.9 & GPSR\,079.602$+$0.506 & \cite{zoone1990} & \\
0933 & ----- & ----- &20:32:36.60 & 41:14:47.87 & 14.4 & RLP\,886  & \cite{reddish1966} & star\\
 ----- & 2494 &----- & 20:32:40.83 & 41:14:29.31 & 1.4 & Cyg\,OB2\,12 &\cite{pacwbcat} & star \\
0941 & 2501 & $+0.6\pm0.1$ &20:32:45.44 & 40:39:37.50 & 1.3 & EM*\,MWC\,349 & \cite{zhang-q2017} & emission-line star\\
0946 & 2514 & $-0.9\pm0.1$ &20:32:55.32 & 40:31:30.99 & 4.6 & 19P\,22 & \cite{wendker1991} &  \\
0953 & 2526 & $-1.1\pm0.1$ &20:33:10.27 & 40:41:16.91 & 1.9 & J203310.31$+$404118.72 & \cite{kryukova2014} & YSO\\
 $\,\,\,\,"$ & 2528 & $"$ & 20:33:11.01 & 40:41:32.23 & 4.8 & 19P 24 & \cite{vollmer2010} & \\
 $\,\,\,\,"$  & 2530 & $"$ & 20:33:11.74 & 40:41:49.35 & 2.1 & J203311.80$+$404151.32 & \cite{kryukova2014} & YSO\\
 ----- & 2536 &----- & 20:33:14.93 & 41:18:50.63 & 1.7 & Cyg\,OB2\,8A & \cite{pacwbcat} & OB stellar system\\
0956 & 2544/2545 & $-0.3\pm0.1$ & 20:33:18.92 & 40:58:37.39 & 14.4 & G079.964$+$00.579 & \cite{anderson2015} & \\  
0957 & 2545 & $-0.4\pm0.1$ &20:33:19.00 & 40:59 05.06 & 13.6 & G079.964$+$00.579 & \cite{anderson2015} & \\
 $\,\,\,\,"$ & ----- &----- & 20:33:19.00 & 40:59 05.06 & 15.5 & BD$+$40 4230 & \cite{reddish1966} & star\\
0963 & 2552/2556/2558 &$-1.2\pm0.1$ & 20:33:23.42 & 41:27:17.60 & 6.6 & 19P\,26 & \cite{vollmer2010} & \\
$\,\,\,\,"$ & $\,\,\,\,"$ & $"$ &20:33:23.42 & 41:27:17.60 & 22.3 & IDX\,114 & \cite{rauw2011} & X-ray star \\  
0965 & 2559 & $-0.8\pm0.3$ &20:33:24.95 & 40:57:28.81 & 8.3 & RLP\,1034 & \cite{reddish1966} & star\\
0966 & 2560  & $-0.5\pm0.3$ &20:33:26.58 & 40:42:33.04 & 1.8 & J203326.44$+$404233.74 & \cite{kryukova2014} & YSO \\
 ----- &  2565 &-----& 20:33:30.74 & 41:35:28.42 & 8.5 & RLP\,283  & \cite{reddish1966} & star\\
0968 & 2566 & $-0.3\pm0.1$ & 20:33:31.97 & 40:41:03.03 & 8.2 & 19P\,28 & \cite{wendker1991} & \\
0976 & 2585 & $-1.4\pm0.1$ & 20:33:47.12 & 40:40:54.64 & 10.6 & J203348.01$+$404051.61 &\cite{kryukova2014} & YSO \\
0977 & 2589 & $-0.9\pm0.1$ & 20:33:52.22 & 41:15:45.11 & 7.7 & AFM2007--990 & \cite{facundo2007} & X-ray star \\ 
  ----- & 2603 &-----& 20:34:06.69 & 41:16:00.86 & 1.0 & J203406.77$+$411600.46 & \cite{kryukova2014} & YSO\\
0985 & 2607 & $-0.3\pm0.2$ & 20:34:10.56 & 41:06:58.71 & 0.7 & IPHASX\,J203410.5$+$410659 & \cite{wright2012} & frEGG\\
0988 & 2610  & $-2.5\pm0.2$ & 20:34:13.85 & 41:08:16.31 & 5.0 & WDDGGHK7 & \cite{isequilla2019} & frEGG \\  
1000 & 2635 & $-1.6\pm0.2$ & 20:34:36.48 & 40:51:59.24 & 5.3 & WDDGGHK4 & \cite{isequilla2019} & frEGG\\
1003 & 2638 & $-1.3\pm0.1$ & 20:34:43.28 & 40:53:15.48 & 2.1 & WDDGGHK3 & \cite{isequilla2019} & frEGG\\
 ------ & 2652 &-----& 20:34:53.33 & 40:53:20.89 & 2.2 &  WDDGGHK2 & \cite{isequilla2019} & frEGG \\  
1006 & 2641 & $-0.8\pm0.3$ & 20:34:45.15 & 41:45:03.2 & 3.2 & GPSR\,079.918$+$0.283 & \cite{zoone1990} & \\
 ----- & 2625 &-----& 20:34:29.54 & 41:31:45.2 & 0.7 & TYC\,3161--1048--1 & \cite{tychocat2000} & star\\
1012 & 2654 & $-0.6\pm0.1$ & 20:34:55.80 & 40:40:46.56 & 1.6 & GPSR\,079.904$+$0.154 & \cite{zoone1990} & \\
 ----- & 2657 &-----& 20:34:57.30 & 40:04:14.94 & 6.5 & TYC\,3157--1182--1 & \cite{tychocat2000} & star \\
1014 & 2659 & $+0.4\pm0.1$ & 20:35:00.28 & 41:34:52.90 & 2.34 & IRAS\,20332$+$4124 & \cite{lux2014} & star-forming region\\
1018 & 2663 & $-0.0\pm0.2$ & 20:35:02.77 & 41:34:51.22 & 10.8 & IRAS\,20332$+$4124 & \cite{lux2014} & star-forming region\\ 
1019 & 2664 &  $-0.1\pm0.2$ & 20:35:03.56 & 41:18:24.12 & 0.4 & 2MASS\,J20350353$+$4118240 & \cite{2MASScat} &  IR source\\
 ----- & 2666 &-----& 20:35:07.84 & 39:59:48.46 & 2.8 & GPSR\,079.380$-$0.284 & \cite{zoone1990} & \\
1020 & 2670 & $+0.9\pm0.1$ & 20:35:16.63 & 40:49:44.66 & 1.0 & GPSR\,080.063$+$0.191 & \cite{zoone1990} &\\
1024 & 2680 &$-1.2\pm0.1$& 20:35:32.37 & 41:44:56.33 & 0.8 & NVSS\,J203532$+$414456 & \cite{NVSS} &\\
1025 & -----  & -----& 20:35:33.17 & 41:06:45.07 & 2.8 & GPSR\,080.322$+$0.319 & \cite{vollmer2010} & \\  
1027 & ----- & -----& 20:35:42.72 & 40:52:51.58 & 1.7 & GPSR\,080.154$+$0.156 & \cite{zoone1990} & \\
1030 & 2689 &$+0.2\pm0.2$& 20:35:47.07 & 41:22:45.01 & 0.4 & WR\,146 & \cite{pacwbcat} & WR system \\
 ----- & 2693 & -----& 20:35:55.32 & 42:18:03.67 & 3.5 & 2034+42A & \cite{clegg1992} & \\
 ----- & 2697 & -----& 20:35:58.48 & 42:17:23.86 & 3.1 & 2034+42B &  \cite{clegg1992} &  \\  
1032 & 2698 &$-0.6\pm0.2$& 20:35:59.52 & 40:54:00.86 & 12.5 & J203558.60$+$405353.85 & \cite{kryukova2014} & YSO\\
$\,\,\,\,''$ & $\,\,\,\,''$ & $''$ & 20:36:01.31 & 40:53:56.79 & 4.4 & J203600.94$+$405358.24 & \cite{kryukova2014} & YSO\\
$\,\,\,\,''$ & 2700 & $''$ & 20:36:01.31 & 40:53:56.79 & 8.1 & J203601.85$+$405351.48 & \cite{kryukova2014} & YSO \\
1034 & 2707 & $+0.2\pm0.2$& 20:36:12.84 & 40:45:43.14 & 2.1 & GPSR\,080.115$+$0.009 & \cite{zoone1990} & \\
1038 & 2721 & $-0.6\pm0.1$ & 20:36:29.72 & 41:20:21.90 & 0.5 & CPR2002--B3 & \cite{comeron2002} & star\\
1039 & 2722 & $-1.2\pm0.1$& 20:36:34.45 & 41:32:22.66 & 10.9 & G080.522$+$00.714 & \cite{zoone1990} & \\
 ----- & 2730 & -----& 20:36:43.72 & 40:21:09.99 & 2.6 & WR\,147 & \cite{pacwbcat} & WR system \\
 ----- & 2770 & -----& 20:37:37.72 & 40:53:52.54 & 1.8 & 080.386$-$0.122  & \cite{garwood1988}  & \\  
 ----- & 2785 & -----& 20:37:58.29 & 40:00:52.85 & 2.9 & 18P\,6 & \cite{vollmer2010} & \\
 ----- & 2793 & -----& 20:38:22.23 & 40:16:16.95 & 2.0 & GPSR\,079.972$-$0.614 & \cite{zoone1990} & \\
\label{counterparts}
\end{longtable}
\end{landscape}

\end{document}